%% file: TMD_power_v2.tex
\documentclass[preprintnumbers,notitlepage,prd,floatfix,nofootinbib,superscriptaddress]{revtex4-1}

\include{settings}

\allowdisplaybreaks[2]

\begin{document}


\preprint{JLAB-THY-20-3159}

\title{Predictive power of transverse-momentum-dependent distributions} 

\author{Manvir Grewal}
\email{manvir.grewal@columbia.edu}
\affiliation{Department of Physics, Columbia University, New York, NY 10027, USA}

\author{Zhong-Bo Kang}
\email{zkang@physics.ucla.edu}
\affiliation{Department of Physics and Astronomy, University of California, Los Angeles, CA 90095, USA}
\affiliation{Mani L. Bhaumik Institute for Theoretical Physics, University of California, Los Angeles, CA 90095, USA}
\affiliation{Center for Frontiers in Nuclear Science, Stony Brook University, Stony Brook, NY 11794, USA}

\author{Jian-Wei Qiu}
\email{jqiu@jlab.org}
\affiliation{Theory Center, Thomas Jefferson National Accelerator Facility, 12000 Jefferson Avenue, Newport News, VA 23606, USA}

\author{Andrea Signori}
\email{asignori@jlab.org}
\affiliation{Dipartimento di Fisica, Universit\`a di Pavia, via Bassi 6, I-27100 Pavia, Italy}
\affiliation{INFN, Sezione di Pavia, via Bassi 6, I-27100 Pavia, Italy}
\affiliation{Theory Center, Thomas Jefferson National Accelerator Facility, 12000 Jefferson Avenue, Newport News, VA 23606, USA}

\begin{abstract}
We investigate the predictive power of transverse-momentum-dependent (TMD) distributions as a function of the light-cone momentum fraction $x$ and the hard scale $Q$ defined by the process. 
We apply the saddle point approximation to the unpolarized quark and gluon transverse momentum distributions and evaluate the position of the saddle point as a function of the kinematics.  
We determine quantitatively that the predictive power for an unpolarized transverse momentum distribution is maximal in the large-$Q$ and small-$x$ region. For cross sections the predictive power of the 
TMD factorization formalism is generally enhanced by considering the convolution of two distributions, and we explicitly consider the case of $Z$ and $H^0$ boson production. 
In the kinematic regions where the predictive power is not maximal, the distributions are sensitive to the non-perturbative hadron structure. Thus, these regions are critical for investigating hadron tomography in a three-dimensional momentum space. 
\end{abstract}

\date{\today}
\maketitle
\tableofcontents

\section{Introduction}
\label{s:intro}

The theoretical study and experimental exploration of the internal structure of nucleons are of fundamental importance to science~\cite{Boer:2011fh,Accardi:2012qut,Lin:2017snn}. In the past decades, we have obtained a detailed knowledge of the so-called collinear parton distribution functions (PDFs). These collinear PDFs describe the distribution of partons inside a fast moving 
nucleon as a function of the nucleon's longitudinal momentum fraction $x$, and thus provide us with a ``one-dimensional'' (1D) picture of how partons are distributed inside the nucleons. 
They are indispensable in the predictions involving high-energy hadrons, such as those at the Large Hadron Collider (LHC), in particular for the inclusive observables with one large momentum transfer, e.g., the total cross section of $W/Z$ and $H^0$ bosons computed in the collinear factorization formalism~\cite{Collins:1989gx}. 

On the other hand, for the observables with more than one observed momentum scale, such as the transverse momentum distribution of $W/Z$ and $H^0$ bosons 
when the transverse momentum is so much smaller than the mass of the observed particle ($q_T\ll Q\sim M_{W/Z,H^0}$), 
a more sophisticated factorization framework, namely the transverse-momentum-dependent (TMD) factorization~\cite{Collins:1984kg,Ji:2004wu,Ji:2004xq,GarciaEchevarria:2011rb,Collins:2011zzd}, is needed. In such a TMD factorization framework, the observables are written in terms of transverse-momentum-dependent PDFs (TMD PDFs), which are usually just called TMDs for simplicity. The TMDs contain not only the aforementioned longitudinal momentum fraction $x$, but also the partonic transverse momentum~$k_T$ with respect to the direction of the 
parent nucleon. Because of this, the TMDs provide us 
the rich information on ``three-dimensional'' (3D) motion of the probed active parton inside the nucleon, often referred to as 3D imaging of the nucleon~\cite{Boer:2011fh,Accardi:2012qut,Aidala:2020mzt}. 

Owing to one of the key defining properties of Quantum Chromodynamics (QCD), the color confinement, we do not see any quarks and gluons in isolation.  It is therefore critically important to have a reliable and controllable {\em matching} between the properties and dynamics of quarks and gluons participating in high energy collisions and the hadrons observed in the detector, which could be achieved by the QCD factorization.  Thus, the investigation of the TMDs and the associated TMD factorization becomes extremely important. On one side, they have a strong interplay with high-energy physics, since the uncertainties of hadronic nature as encoded in the TMDs are among the largest ones 
that dominate the systematic theoretical uncertainties for the QCD calculations of key observables, 
which could impact our ability to explore the possible scenarios of Beyond the Standard Model physics. 
On the other hand, a good knowledge of TMDs is essential to map out 
the nucleon's 3D partonic structure, namely to understand the confined motion of quarks and gluons inside a bound hadron. 
This is particularly true in light of the rapid progress towards realizing a US-based Electron-Ion Collider (EIC), a machine aiming at investigating the multidimensional structure of hadrons and nuclei. 

%
\begin{figure}[h!]
\centering
\includegraphics[width=0.35\textwidth]{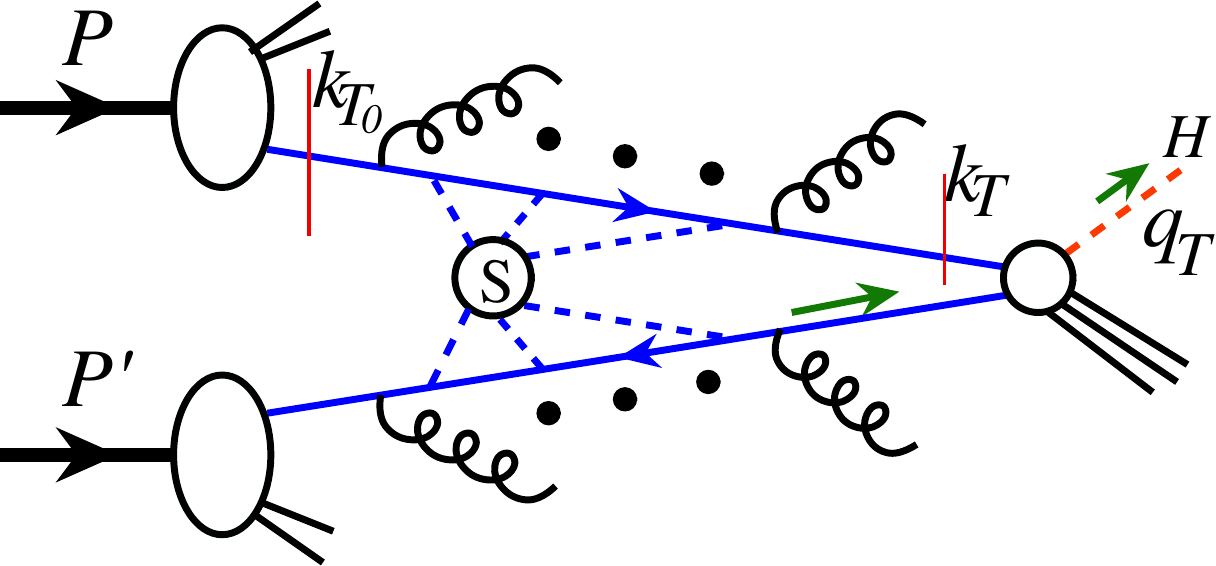}
\caption{Sketch for the Drell-Yan type of heavy boson production with the parton shower.}
\label{fig:dy}
\end{figure}

In both frontiers, one of the most important questions to address would be to understand in which kinematic regions the perturbative QCD-based formalism of TMD factorization is most predictive, from the point of view of a controllable perturbative computation. 
To address this question, it is important to recognize that the probed transverse momentum ($k_T$) of the active parton in the hard collisions is not the same as the transverse momentum of the same parton inside a bound hadron, sometimes referred to as the intrinsic $k_{T_0}$ and as shown in Fig.~\ref{fig:dy} for a generic Drell-Yan type hard collision. 
With the hard collision and the large momentum transfer, a large amount of parton shower is developed during the collision, making the $k_T$ of the probed active parton different from the intrinsic transverse momentum $k_{T_0}$ associated with the confined motion inside the bound hadron. 
The difference between the $k_T$ and $k_{T_0}$ depends on the hard scale of the collision, $Q$, as well as the phase space available for the shower or the total collision energy $\sqrt{s}$. 
The observed $Q$ and $\sqrt{s}$ determine the momentum fraction $x$ of the active parton participating in the hard collision. The smaller $x$ is, the larger the phase space available for the shower is.  The difference between the $k_T$ and $k_{T_0}$ is encoded in the QCD evolution of the TMDs in terms of the TMD factorization. 
As demonstrated quantitatively in this paper, the QCD evolution of the $k_T$-dependence could be dominated by the logarithmic and perturbatively calculable part of the parton shower, leading to a better predictive power. Moreover, the intrinsically non-perturbative TMDs could be further factorized into the non-perturbative 1D collinear PDFs convoluted with calculable contributions from the parton shower.  On the other hand, if the evolution of the $k_T$-dependence is dominated by the non-perturbative dynamics of the parton shower, the TMDs and the corresponding observables will be more sensitive to the non-perturbative physics. 
The detailed and quantitative study presented in this paper will help us figure out in which regions these non-perturbative contributions play a significant role and where the experimental data are most ideal in order to constrain the non-perturbative component of the TMDs.

TMD factorization and evolution have been extensively studied in the literature~\cite{Collins:1984kg,Collins:1981uw,Ji:2004wu,Ji:2004xq,Collins:2011zzd,GarciaEchevarria:2011rb,Rogers:2015sqa,Angeles-Martinez:2015sea,Collins:2017oxh}, together with the matching to collinear factorization~\cite{Collins:1984kg,Arnold:1990yk,Nadolsky:2002jr,Berger:2004cc,Stewart:2013faa,Collins:2016hqq,Echevarria:2018qyi}, the generalized universality properties~\cite{Collins:2002kn,Boer:2003cm,Collins:2004nx,Bomhof:2007xt,Kang:2009bp,Gamberg:2010uw,Boer:2010ya,Buffing:2012sz}, and the impact on high-energy physics~\cite{Berger:2002ut,Bacchetta:2018lna,Bozzi:2019vnl}. Much of the efforts in TMD phenomenology is devoted to the understanding of the role and the size of the non-perturbative corrections in different kinematic domains~\cite{Angeles-Martinez:2015sea,Bacchetta:2017gcc,Scimemi:2017etj,Bertone:2019nxa,Bacchetta:2019sam,Scimemi:2019cmh}. The study of the kinematic dependence originates from the work of Parisi and Petronzio~\cite{Parisi:1979se} and Collins, Soper, and Sterman~\cite{Collins:1984kg}, which focused on the value of the hard scale of the process compared to the infrared scale of QCD ($\Lambda_{\rm QCD}$). 
More recently, it has been shown at the level of the cross section~\cite{Qiu:2000hf,Berger:2002ut,Berger:2003pd,Berger:2004cc} that also the light-cone momentum fraction 
$x$, which is effectively a measure of available phase space for the parton shower, could play an important role 
in determining the relevance of the non-perturbative corrections. 
In this article we extend those arguments to the context of the modern TMD factorization formalism~\cite{Collins:2011zzd}, linking the predictive power of the TMDs to their double scale evolution, i.e. the ultraviolet and rapidity renormalization scales to be defined below. Our detailed study shows that for TMDs with the large hard-scale $Q$ and the small momentum fraction $x$, the non-perturbative contribution plays a less important role and thus they have the most predictive power. On the contrary, TMDs with the small hard-scale $Q$ and the large momentum fraction $x$ receive significant non-perturbative contributions, and are better suited for constraining non-perturbative parameters in the TMDs. 

The paper is organized as follows. 
In Sec.~\ref{s:bT_regions} we present the structure of a TMD PDF in the coordinate $b_T$ space, which is conjugate to the transverse momentum $k_T$. We separate the small and large $b_T$ regions, and derive a functional form that extrapolates the physics from the small to the large $b_T$ region. In Sec.~\ref{s:saddle} we apply the saddle-point method to the TMD PDF, and we determine the position of the saddle point as a function of the kinematics studying the structure of the double-scale evolution of the distribution. In Sec.~\ref{s:NP_relevance} we analyze the predictive power of the quark and gluon TMDs 
and comment on the relevance of the large $b_T$ region and its components. 
In Sec.~\ref{s:cross_sections} we study the transverse momentum distributions for $Z$ boson production and $H^0$ boson production in $pp$ collisions. They are sensitive to quark and gluon 
TMDs, respectively. We close the paper in Sec.~\ref{s:conclusions} and comment on the advantages presented by the complementary kinematic regions accessed by different experiments 
and possibilities to learn and control the non-perturbative evolution of TMDs.

\section{TMDs from small to large $b_T$ region}
\label{s:bT_regions}
Our main focus in this paper is on the unpolarized TMD PDF for a parton with specific flavor $a$, 
\bea
F_a(x, k_T^2; \mu, \zeta)\,, 
\eea
which carries the collinear momentum fraction $x$ of the parent hadron and has a transverse momentum $k_T$ with respect to the hadron's momentum. On the other hand, $\mu$ and $\zeta$ are the ultraviolet (UV) renormalization and rapidity 
regularization scales, respectively. As we will discuss in Sec.~\ref{s:cross_sections}, these TMDs are indispensable in describing e.g., the transverse momentum distribution of a vector boson $Z$ and $H^0$ boson production in the low transverse momentum region $q_T\ll M_{Z,H^0}$, 
and carry rich information on the parton's confined motion in a bound hadron, which is a fundamental emergent property of the QCD dynamics. As shown in Fig.~\ref{fig:dy}, the $k_T$-dependence of the TMD PDF probed at the hard collision is a combination of parton's intrinsic $k_{T_0}$ and the amount of $k_T$ generated by the parton shower. 
Since each radiation from the parton shower could be soft and non-perturbative, and convoluted with additional radiation before and after, 
it could be advantageous to study the TMDs in their Fourier transformed form in the position or $b_T$-space, defined as~\cite{Collins:1984kg}
\begin{align}
F_a(x, b_T^2; \mu, \zeta) = \int d^2 {\bm k}_T \,e^{i{\bm k}_T\cdot {\bm b}_T}F_a(x, k_T^2; \mu, \zeta)\,. 
\end{align} 
When $b_T$ is small, much less than $1/\Lambda_{\rm QCD}$, the QCD evolution (or scale dependence) of the TMDs' $b_T$-dependence is perturbatively calculable. Otherwise, the QCD evolution is non-perturbative.  
Once we understand the TMD PDF in the $b_T$-space, we then Fourier transform it back into the momentum space:
\bea
F_a(x, k_T^2; \mu, \zeta) =& \int \frac{d^2{\bm b}_T}{(2\pi)^2} \,e^{-i{\bm k}_T\cdot {\bm b}_T} F_a(x, b_T^2; \mu, \zeta) 
\nonumber \\
=& \frac{1}{2\pi} \int_0^{\infty} db_T \, b_T J_0(k_Tb_T) F_a(x, b_T^2; \mu, \zeta)\, .
\label{eq:F-b-space}
\eea
The zero-th order Bessel function $J_0$ emerges from the angular part of the integral and the absence of any dependence on the azimuthal angle of the transverse momentum ${\bm k}_T$ in the unpolarized case. 
The above Fourier transform would require the information of the 
$F_a(x, b_T^2; \mu, \zeta)$ for the entire $b_T\in [0,\infty)$ region. 
If the Fourier transform is dominated by the information of TMDs at small $b_T$, we will have a good predictive power for $F_a(x, k_T^2; \mu, \zeta)$ in all relevant $k_T$ region, modulo the knowledge of the standard collinear PDFs, as we demonstrate below. 
On the other hand, if the Fourier transform is sensitive to the large $b_T$ region, $F_a(x, b_T^2; \mu, \zeta)$ will be sensitive to non-perturbative physics since the evolution kernels for the scale-dependence of the TMDs at large $b_T$ are non-perturbative. 

Below we first review the behavior and evolution of TMDs in the small-$b_T$ region, and we then study how one can extrapolate the TMD to the large-$b_T$ region by extending the work of Ref.~\cite{Qiu:2000hf}. By further studying the behavior of the TMDs in the $b_T$-space through a saddle-point approximation, we 
explore quantitatively in which region the TMDs have the most predictive power. 

\subsection{TMDs in the small-$b_T$ region}
The QCD evolution equations of the TMDs take the following form
\bea
\frac{d\ln F_a(x, b_T^2; \mu, \zeta)}{d\ln \zeta} &= - D\left(b_T\mu, \alpha_s(\mu)\right)\,,
\\
\frac{d\ln F_a(x, b_T^2; \mu, \zeta)}{d\ln \mu} &= \gamma_F\left(\alpha_s(\mu), \frac{\zeta}{\mu^2}\right)\,,
\\
\frac{d\, D\left(b_T\mu, \alpha_s(\mu)\right)}{d\ln\mu^2} &= \frac{1}{2} \gamma_K\left(\alpha_s(\mu)\right)\,,
\label{eq:K}
\\
\frac{d\,\gamma_F\left(\alpha_s(\mu), \frac{\zeta}{\mu^2}\right)}{d\ln\zeta} &= - \gamma_K\left(\alpha_s(\mu)\right)\,.
\label{eq:gamma_F}
\eea
Here the first three equations are well-known and can be found in the literature, see e.g.~in Ref.~\cite{Collins:2011zzd}, where $D\left(b_T\mu, \alpha_s(\mu)\right)$ is called the Collins-Soper evolution kernel\footnote{Note that we use a slightly different notation and normalization with respect to Ref.~\cite{Collins:2011zzd}.}, and $\gamma_F\left(\alpha_s(\mu), \frac{\zeta}{\mu^2}\right)$ is the anomalous dimension of the operator defining the TMD PDF. The last equation is obtained from the fact that the differential order in $\zeta$ and in $\mu$ for $F_a(x, b_T^2; \mu, \zeta)$ is interchangeable, i.e.,
\bea
\frac{d}{d\ln \zeta}\frac{d}{d\ln\mu}\ln F_a(x, b_T^2; \mu, \zeta) = \frac{d}{d\ln\mu} \frac{d}{d\ln \zeta} \ln F_a(x, b_T^2; \mu, \zeta)\, ,
\eea
so long as $F_a(x, b_T^2; \mu, \zeta)$ are differentiable in both $\mu$ and $\zeta$ in the kinematic regime that we are interested in.

In the perturbative region where $1/b_T\gg \Lambda_{\rm QCD}$, one can compute all the evolution kernels in the above evolution equations. For example, for a quark TMD PDF with $a=q$, we have
\bea
D\left(b_T\mu, \alpha_s(\mu)\right) &= C_F \sum_{n=1}\left(\frac{\alpha_s}{4\pi}\right)^n \sum_{k=0}^n d^{(n,k)}\ln^k\left(\frac{\mu^2}{\mu_b^2}\right),
\\
\gamma_K\left(\alpha_s(\mu)\right) &= \Gamma_{\rm cusp} \left(\alpha_s(\mu)\right),
\\
\gamma_F\left(\alpha_s(\mu), \frac{\zeta}{\mu^2}\right) &= \Gamma_{\rm cusp} \left(\alpha_s(\mu)\right) \ln\left(\frac{\mu^2}{\zeta}\right) + \gamma\left(\alpha_s(\mu)\right),
\eea											
where we define $\mu_b = c/b_T$ with $c=2e^{-\gamma_E}$ and $\gamma_E=0.577$ the Euler constant. $\Gamma_{\rm cusp} \left(\alpha_s(\mu)\right)$ and $\gamma\left(\alpha_s(\mu)\right)$ are the cusp and non-cusp anomalous dimensions, respectively. They generally have the expansion $\Gamma_{\rm cusp} \left(\alpha_s(\mu)\right)=\sum_{n=1}\Gamma_{n-1}\left(\frac{\alpha_s}{4\pi}\right)^n$, likewise for the non-cusp. For a quark TMD PDF, one has $\Gamma_0 = 4C_F,~\gamma_0 = 6C_F$, etc. At the same time, we have $d^{(1,1)} = 2\Gamma_0,~d^{(1,0)} = 0$, etc. Thus at the first non-trivial order, we have
\bea
D\left(b_T\mu, \alpha_s(\mu)\right) &= \frac{\alpha_s}{2\pi} C_F \ln\left(\frac{\mu^2}{\mu_b^2}\right),
\\
\gamma_K\left(\alpha_s(\mu)\right) &= \frac{\alpha_s}{\pi} C_F,
\\
\gamma_F\left(\alpha_s(\mu), \frac{\zeta}{\mu^2}\right) &= \frac{\alpha_s}{\pi} C_F \left[ \ln\left(\frac{\mu^2}{\zeta}\right) + \frac{3}{2}\right]. 
\eea
The higher-order expressions, and the expressions for gluon TMD PDF can be found in e.g. Ref.~\cite{Echevarria:2016scs}. See also Refs.~\cite{Luo:2019hmp,Luo:2019bmw,Luo:2019szz}. 

Solving the evolution equation, one can obtain the evolved TMD PDF as
\begin{equation}
\label{e:pert_TMDPDF_bT}
F_a(x,b_T^2;\mu,\zeta) = F_a(x,b_T^2;\mu_0,\zeta_0)\, 
\exp \bigg[ \int_{\mu_0}^{\mu} \frac{d\mu'}{\mu'}  \gamma_F \left( \alpha_s(\mu'), \frac{\zeta}{\mu'^2} \right) \bigg]\ 
\bigg( \frac{\zeta}{\zeta_0} \bigg)^{-D\left(b_T\mu_0, \alpha_s(\mu_0)\right)} \ ,
\end{equation}
where $\mu_0$ and $\zeta_0$ are the initial values for the renormalization scales. Integrating Eq.~\eqref{eq:gamma_F} from $\mu^2$ to $\zeta$, one obtains
\bea
\gamma_F \left( \alpha_s(\mu), \frac{\zeta}{\mu^2} \right) =  - \ln\left(\frac{\zeta}{\mu^2}\right) \gamma_K\left(\alpha_s(\mu)\right) + \gamma_F \left( \alpha_s(\mu), 1 \right),
\eea
and thus we have
\bea
F_a(x,b_T^2;\mu,\zeta) =& F_a(x,b_T^2;\mu_0,\zeta_0)
\nonumber\\
&\times\exp \Bigg\{ -\bigg[\int_{\mu_0}^{\mu} \frac{d\mu'}{\mu'}   \left( \ln\left(\frac{\zeta}{\mu'^2}\right) \gamma_K\left(\alpha_s(\mu')\right) -\gamma_F \left( \alpha_s(\mu'), 1 \right) \right) 
 + D\left(b_T\mu_0, \alpha_s(\mu_0)\right) \ln\left( \frac{\zeta}{\zeta_0}\right) \bigg]\Bigg\}\,.
\eea
Finally when both $\mu_0$ and $\zeta_0$ are in the perturbative region, the TMD PDF $F_a$ at the input scales $\mu_0$ and $\zeta_0$ can be re-factorized onto collinear PDFs $f_b(x, \mu_0)$ via an operator product expansion (OPE) at low $b_T$:
\begin{equation}
\label{e:input_TMDPDF}
F_a(x,b_T^2;\mu_0,\zeta_0) = 
\sum_b C_{a/b}(x,b_T^2,\mu_0,\zeta_0) \otimes f_b(x,\mu_0) =  
\sum_b \int_x^1  \frac{d\hat x}{\hat x} C_{a/b}\bigg(\hat x,b_T^2,\mu_0,\zeta_0\bigg) f_b\left(\frac{x}{\hat x},\mu_0\right) \, .
\end{equation}

In practice, one typically chooses the following input values for $\mu_0$ and $\zeta_0$,
\bea
\zeta_0 = \mu_0^2  = \mu_b^2,
\eea
to eliminate the logarithms in the coefficient functions $C_{a/b}\bigg(\hat x,b_T^2,\mu_0,\zeta_0\bigg)$. At the same time, one usually chooses $\mu$ and $\zeta$ to be associated with the hard scale $Q$, such as the invariant mass of the lepton pair in the Drell-Yan process, $pp\to[\gamma^*\to] \ell^+\ell^- + X$, 
\bea
\zeta = \mu^2 = Q^2.
\eea
Thus in the usual phenomenology we write the perturbative TMD PDF in Eq.~\eqref{e:pert_TMDPDF_bT} in the following form 
\bea
F_a(x,b_T^2; Q, Q^2) =& F_a(x,b_T^2;\mu_b,\mu_b^2)
\nonumber\\
&\times \exp \Bigg\{ - \bigg[ \int_{\mu_b}^{Q} \frac{d\mu'}{\mu'}   \left( \ln\left(\frac{Q^2}{\mu'^2}\right) \gamma_K\left(\alpha_s(\mu')\right) - \gamma_F \left( \alpha_s(\mu'), 1 \right) \right)  + D\left(c, \alpha_s(\mu_b)\right) \ln\left( \frac{Q^2}{\mu_b^2}\right) \bigg] \Bigg\}\,,
\label{e:F_pert}
\eea
where $F_a(x,b_T^2;\mu_b,\mu_b^2)$ will be obtained through Eq.~\eqref{e:input_TMDPDF}. 

\subsection{Extrapolation to the large-$b_T$ region}
From the discussion of the previous section, we gain the following information. In the kinematic region where $Q$ is large ($Q\gg \Lambda_{\rm QCD}$) and for the small-$b_T$ region, one can rely on the perturbative result in Eq.~\eqref{e:F_pert} to obtain the information for the TMD PDF $F_a(x,b_T^2; Q, Q^2)$. However, for the large-$b_T$ region, non-perturbative physics kicks in and the perturbative result is no longer reliable. 
Several proposals have been introduced to extrapolate the TMD PDF at small-$b_T$ into the large-$b_T$ region~\cite{Collins:1984kg}. 
In this paper, we follow the spirit of Ref.~\cite{Qiu:2000hf} to keep the TMD PDF at small-$b_T$ unchanged while we derive a functional form to extrapolate the perturbative result in the small-$b_T$ region to the large-$b_T$ region. Such an extrapolation would preserve the predictive power of the perturbative calculations in the small-$b_T$ region, which is not affected by the extrapolation at large $b_T$, and at the same time, it would provide a physically motivated functional form for the large-$b_T$ region.  
In other words, we write the TMD PDF in the $b_T$-space as
\begin{equation}
\label{e:piecewise_TMD}
F_a(x,b_T^2; Q, Q^2) = 
\begin{cases} 
     F_a^{\rm \tiny{OPE}}(x,b_T^2; Q, Q^2) & \qquad b_T \leq \bmax\, , \\
     F_a^{\rm \tiny{OPE}}(x,b_{\rm max}^2; Q, Q^2) R_a^{\rm NP}(x, b_T, Q; b_{\rm max}) & \qquad b_T > \bmax \, ,
\end{cases} 
\end{equation}
where the parameter $\bmax$ is the largest value of $b_T$ at which the perturbative expression for the TMD PDF is trusted 
(like the input scale at which the DGLAP evolution starts for the 1D PDFs). 
We choose a rather conservative value for $\bmax = 0.5$ GeV$^{-1}$ throughout this paper. Accordingly, for $b_T \leq \bmax$, $F_a^{\rm \tiny{OPE}}(x,b_T^2; Q, Q^2)$ is just the perturbative expression given in Eq.~\eqref{e:F_pert}\footnote{In this paper we do not consider the corrections needed for the proper treatment of the region at an extremely small $b_T$~\cite{Parisi:1979se,Bozzi:2005wk,Boer:2014tka,Collins:2016hqq,Bacchetta:2017gcc}, which is phenomenologically relevant typically at energies lower than the ones considered in our analyses.}. Here we use the superscript ``OPE'' to remind that Eq.~\eqref{e:F_pert} is connected with the collinear PDFs through an OPE, see Eq.~\eqref{e:input_TMDPDF}. For $b_T > \bmax$, instead, the non-perturbative correction factor $R_a^{\rm NP}$ tames the behavior of $F_a$ when the perturbative calculation is not to be trusted. To maintain the continuity of the TMD PDF at $b_T=\bmax$, the extrapolation function $R_a^{\rm NP}$ should satisfy
\bea
R_a^{\rm NP}(x, b_T = b_{\rm max}, Q; b_{\rm max}) = 1\,. 
\label{e:norm}
\eea

To derive a functional form for $R_a^{\rm NP}$, we take into account the power correction in the evolution kernel~\cite{Qiu:2000hf}. The Collins-Soper kernel $D\left(b_T\mu, \alpha_s(\mu)\right)$ has an explicit $b_T$-dependence. When $b_T> \bmax$, we add a power correction into its evolution equation as follows~\footnote{The type of power correction in the form of $1/Q^2$ is studied in Ref.~\cite{Balitsky:2017gis}. 
}
\bea
\frac{d\, D\left(b_T\mu, \alpha_s(\mu)\right)}{d\ln\mu^2} &= \frac{1}{2} \left[\gamma_K\left(\alpha_s(\mu)\right) + \frac{1}{\mu^2}{\overline{\gamma}_K}\right],
\label{eq:K-evo}
\eea
where $\overline{\gamma}_K$ is an unknown parameter that characterizes the typical size of the higher-twist operator. Such a power correction to the evolution equation is also referred to as ``dynamical power correction'' in~\cite{Qiu:2000hf} and we will continue to use this terminology. For consistency within the TMD evolution equations Eqs.~\eqref{eq:K} and \eqref{eq:gamma_F}, one would also have 
\bea
\frac{d\,\gamma_F\left(\alpha_s(\mu), \frac{\zeta}{\mu^2}\right)}{d\ln\zeta} &= - \gamma_K\left(\alpha_s(\mu)\right) -  \frac{1}{\mu^2}{\overline{\gamma}_K}.
\eea
With the modified evolution equations, choosing initial scales for the evolution $\zeta_0 = \mu_0^2=\mu_{\bmax}^2$ and final scales $\zeta = \mu^2 = Q^2$, one would obtain
\bea
F_a(x,b_T^2; Q,Q^2) = F_a(x,b_T^2; \mu_{b_{\rm max}}, \mu_{b_{\rm max}}^2)\,
\exp \Bigg\{ &\int_{\mu_{b_{\rm max}}}^{Q} \frac{d\mu'}{\mu'}   \left[\gamma_F \left( \alpha_s(\mu'), 1 \right) - \ln\left(\frac{Q^2}{\mu'^2}\right) \left(\gamma_K\left(\alpha_s(\mu') \right) +\frac{1}{\mu'^2}\overline{\gamma}_K\right) \right]
\nonumber\\
&- D\left(b_T \mu_{b_{\rm max}}, \alpha_s(\mu_{b_{\rm max}})\right)\ln\left(\frac{Q^2}{\mu_{b_{\rm max}}^2}\right)\Bigg\}\,,
\label{e:F-bt}
\eea
where the input scale $\mu_{b_{\rm max}} = c/\bmax$. Setting $b_T = \bmax$ in the above equation, we would obtain
\bea
F_a(x,b_{\rm max}^2; Q,Q^2) = F_a(x,b_{\rm max}^2; \mu_{b_{\rm max}}, \mu_{b_{\rm max}}^2)\,
\exp \Bigg\{ &\int_{\mu_{b_{\rm max}}}^{Q} \frac{d\mu'}{\mu'}   \left[\gamma_F \left( \alpha_s(\mu'), 1 \right) - \ln\left(\frac{Q^2}{\mu'^2}\right) \left(\gamma_K\left(\alpha_s(\mu') \right) +\frac{1}{\mu'^2}\overline{\gamma}_K\right) \right]
\nonumber\\
&- D\left(c, \alpha_s(\mu_{b_{\rm max}})\right)\ln\left(\frac{Q^2}{\mu_{b_{\rm max}}^2}\right)\Bigg\}\,,
\label{e:F-bmax}
\eea
where we have used $b_{\rm max} \mu_{b_{\rm max}} = c$. By comparing Eqs.~\eqref{e:F-bt} and \eqref{e:F-bmax}, we find 
\bea
F_a(x,b_T^2; Q,Q^2) = F_a(x,b_{\rm max}^2; Q,Q^2) R_a^{\rm NP}(x, b_T, Q; b_{\rm max}),
\eea
with the extrapolation function $R_a^{\rm NP}$ given by
\bea
R_a^{\rm NP}(x, b_T, Q; b_{\rm max}) = \frac{F_a(x,b_T^2; \mu_{b_{\rm max}}, \mu_{b_{\rm max}}^2)}{F_a(x,b_{\rm max}^2; \mu_{b_{\rm max}}, \mu_{b_{\rm max}}^2)} \exp \bigg\{ -\ln\left(\frac{Q^2}{\mu_{b_{\rm max}}^2}\right)  \big[D\left(b_T \mu_{b_{\rm max}}, \alpha_s(\mu_{b_{\rm max}})\right) - D\left(c, \alpha_s(\mu_{b_{\rm max}})\right) \big] \bigg\}\,.
\label{e:RNP-first}
\eea

In order to find a reasonable functional form for $R_a^{\rm NP}$, we now need to figure out the following two factors:
\bea
\frac{F_a(x,b_T^2; \mu_{b_{\rm max}}, \mu_{b_{\rm max}}^2)}{F_a(x,b_{\rm max}^2; \mu_{b_{\rm max}}, \mu_{b_{\rm max}}^2)},
\qquad
\big[D\left(b_T \mu_{b_{\rm max}}, \alpha_s(\mu_{b_{\rm max}})\right) - D\left(c, \alpha_s(\mu_{b_{\rm max}})\right)\big].
\label{eq:two}
\eea
For the second factor, we turn to the modified evolution equation for $D\left(b_T \mu_b, \alpha_s(\mu_b)\right)$ in Eq.~\eqref{eq:K-evo}. To proceed, we integrate $\mu^2$ from $\mu_b^2$ to $\mu_{b_{\rm max}}^2$ and obtain
\bea
D\left(b_T \mu_{b_{\rm max}}, \alpha_s(\mu_{b_{\rm max}})\right) - D\left(b_T \mu_b, \alpha_s(\mu_b)\right)
=&\int_{\mu_b^2}^{\mu_{b_{\rm max}}^2} \frac{d\mu^2}{\mu^2} \frac{1}{2} \left[\gamma_K\left(\alpha_s(\mu)\right) + \frac{1}{\mu^2}{\overline{\gamma}_K}\right] 
\nonumber\\
=& \frac{\gamma}{2\alpha (c^2)^\alpha} \left(\left(b_T^2\right)^\alpha - \left(b_{\rm max}^2\right)^\alpha \right) + \frac{\bar\gamma_K}{2c^2} \left(b_T^2 - b_{\rm max}^2\right)
\nonumber \\
\equiv & g_1 \left(\left(b_T^2\right)^\alpha - \left(b_{\rm max}^2\right)^\alpha \right) + g_2 \left(b_T^2 - b_{\rm max}^2\right)\,.
\eea
To obtain the second line on the right-hand side, we approximate the $\mu$-dependence of $\gamma_K(\alpha_s(\mu))\approx \gamma (\mu^2)^{-\alpha}$ with parameters $\gamma$ and $\alpha$~\cite{Qiu:2000hf}. We further define the prefactors on the second line to be parameters $g_1$ and $g_2$. Realizing $b_T \mu_b = c$, we thus obtain
\bea
D\left(b_T \mu_{b_{\rm max}}, \alpha_s(\mu_{b_{\rm max}})\right) - D\left(c, \alpha_s(\mu_{b_{\rm max}})\right) =& D\left(c, \alpha_s(\mu_b)\right) - D\left(c, \alpha_s(\mu_{b_{\rm max}})\right) 
+ g_1 \left( b_T^{2\alpha} - b_{\rm max}^{2\alpha} \right) 
+ g_2 \left(b_T^2 - b_{\rm max}^2\right)\,.
\eea
Note that 
the term 
$D\left(c, \alpha_s(\mu_b)\right) - D\left(c, \alpha_s(\mu_{b_{\rm max}})\right)$  
on the right-hand side  
depends only on $b_T$ and $\bmax$ through the coupling constant $\alpha_s$, and thus such a term can be combined with 
the one proportional to $g_1$ 
(given its connection to the coupling constant), treating $g_1$ and $\alpha$ as fitting parameters. 

For the first factor in Eq.~\eqref{eq:two}, we realize that at the input scale $\mu_{\bmax}$, one usually mimics the $b_T$-dependence of the TMD PDF $F_a(x,b_T^2; \mu_{b_{\rm max}}, \mu_{b_{\rm max}}^2)$ to have a Gaussian form, see e.g. Refs.~\cite{Anselmino:2007fs,Anselmino:2008sga,Signori:2013mda,Anselmino:2013lza,Echevarria:2014xaa},
\bea
F_a(x,b_T^2; \mu_{b_{\rm max}}, \mu_{b_{\rm max}}^2) \approx f_a(x, \mu_{\bmax}) \exp\left[ - \overline{g}_2 b_T^2\right]\,,
\eea 
which describes the intrinsic transverse momentum of the partons. With such an approximation, we thus obtain the ratio of TMD PDF at the input scale $\mu_{\bmax}$ in Eq.~\eqref{e:RNP-first} as
\bea
\frac{F_a(x,b_T^2; \mu_{b_{\rm max}}, \mu_{b_{\rm max}}^2)}{F_a(x,b_{\rm max}^2, \mu_{b_{\rm max}}, \mu_{b_{\rm max}}^2)} \approx \exp\left[ - \overline{g}_2 \left(b_T^2 - \bmax^2\right)\right]\,,
\eea
Combining all the above factors, we obtain the following form for the extrapolation function
\bea
R_a^{\rm NP}(x, b_T, Q; b_{\rm max}) = \exp \Bigg\{ -\ln\left(\frac{Q^2}{\mu_{b_{\rm max}}^2}\right) \left[g_1 \left(\left(b_T^2\right)^\alpha - \left(b_{\rm max}^2\right)^\alpha \right) + g_2 \left(b_T^2 - b_{\rm max}^2\right)\right] - \overline{g}_2 \left(b_T^2- \bmax^2\right) \Bigg\}\,.
\label{eq:R-np}
\eea
Such a derivation is motivated by the work presented in Ref.~\cite{Qiu:2000hf}. Our derivation is for an individual TMD PDF, while Ref.~\cite{Qiu:2000hf} is for the Drell-Yan differential cross section. This new derivation is based on modern TMD evolution for the TMD PDF, which makes the derivation more transparent and more straightforward. 

Our derived extrapolation function $R_a^{\rm NP}$ automatically satisfies the normalization condition in Eq.~\eqref{e:norm}, i.e.~$R_a^{\rm NP}=1$ at $b_T=\bmax$. Besides $\bmax = 0.5$ GeV$^{-1}$ we have chosen beforehand, it consists of four parameters $\alpha,~g_1,~g_2$, and ${\bar g}_2$. While $g_2$ controls the size of the dynamical power correction, ${\bar g}_2$ mimics the intrinsic transverse momentum, which is also referred to as ``intrinsic power correction'' in~\cite{Qiu:2000hf}. These two parameters are non-perturbative in nature and generally have to be determined from fits to the experimental data. As we will emphasize below, we require $F_a(x,b_T^2; Q, Q^2)$ to be smooth at $b_T=\bmax$, in order to determine the other two parameters $g_1$ and $\alpha$ in the extrapolation function $R_a^{\rm NP}$. Specifically, we require the first and second order derivatives of $F_a(x,b_T^2; Q, Q^2)$ to be continuous at $b_T=\bmax$. With these two conditions, $g_1$ and $\alpha$ can be fixed.
Accordingly, the $R_a^{\rm NP}$ function acquires an implicit $x$-dependence (equivalent to a $\sqrt{s}$-dependence) through these two parameters.

\section{Saddle point approximation of a TMD PDF} 
\label{s:saddle}
Once we have the full $b_T$-dependence of a TMD PDF from the extrapolation method discussed in the previous section, we will be able to compute the TMD PDF in the momentum space through the Fourier transformation:
\bea
F_a(x, k_T^2; Q, Q^2) = \frac{1}{2\pi} \int_0^{\infty} db_T \, b_T J_0(k_Tb_T) F_a(x, b_T^2; Q, Q^2)\,,
\label{eq:F-b-space-Q}
\eea
where we have set $\zeta=\mu^2=Q^2$ in Eq.~\eqref{eq:F-b-space}. Obviously if $F_a(x, b_T^2; Q, Q^2)$ in the $b_T$-space is dominated by the small-$b_T$ behavior, the integration on the right-hand side, and thus $F_a(x, k_T^2; Q, Q^2)$ in the $k_T$-space, will be mainly controlled by the perturbative physics. On the contrary, if 
$F_a(x, b_T^2; Q, Q^2)$ is very sensitive to the large-$b_T$ behavior, the non-perturbative physics will play an important role in the behavior of the TMD PDF $F_a(x, k_T^2; Q, Q^2)$ in the momentum space. Understanding the TMD PDF $F_a(x, k_T^2; Q, Q^2)$ in the momentum $k_T$-space, i.e., whether it is more dominated by perturbative (small-$b_T$) or non-perturbative (large-$b_T$) physics, is very important in order to investigate the predictive power of the TMD PDF and of the TMD differential cross sections, which are based on these TMD PDFs. This is the main goal of this and the next sections. 

Following Refs.~\cite{Collins:1984kg,Qiu:2000hf}, we use the saddle-point method to pinpoint if and how the integration on the right-hand side of Eq.~\eqref{eq:F-b-space-Q} is dominated by the small-$b_T$ region. The saddle-point approximation, or the method of steepest descent, is often used to approximate the integral when the integrand has the form of $e^{-c\, S(b_T)}$, where $c$ is a constant and $S$ a smooth function of $b_T$. As the negative exponential function is rapidly decreasing, one only needs to look at the contribution from where the exponent is at its minimum. Since the TMD PDF in $b_T$-space follows such a form, see Eqs.~\eqref{e:F_pert} and \eqref{eq:R-np}, it is natural to apply the saddle-point approximation to analyze the TMD PDF. We mainly concentrate on the case where $k_T=0$. In such a case, $J_0(k_Tb_T)=1$ and no oscillations are present.  When $k_T>0$, the Bessel function $J_0(k_Tb_T)$ further suppresses the large-$b_T$ region of the $b_T$ integration, and our analysis will be further improved. At $k_T=0$, we have
\bea
F_a(x, k_T^2=0; Q, Q^2) = \frac{1}{2\pi} \int_0^{\infty} db_T \, b_T  F_a(x, b_T^2; Q, Q^2) 
=\frac{1}{4\pi} \int_{-\infty}^\infty d\left(\ln b_T^2\right)  \exp\left[ \ln \left(b_T^2\, F_a(x, b_T^2; Q, Q^2) \right)\right]\,,
\label{eq:integral}
\eea
and thus the integral is dominated by a saddle point at $b_T^{sp}$, which is determined by~\cite{Collins:1984kg}
\bea
\frac{d}{db_T} \bigg\{ \ln \bigg[ b_T^2\, F_a(x,b_T^2; Q, Q^2) \bigg] \bigg\}_{\begin{subarray}{l} b_T = b_T^{sp} \\  \end{subarray}} = 0\,.
\label{eq:saddle-point}
\eea
In the following, we will study in details the kinematic dependence of the saddle point $b_T^{sp}$, in particular the most relevant $x$ and $Q$ dependence: 
\bea 
b_T^{sp} \equiv b_T^{sp}(x, Q).
\eea
The approximation relates the integral over $b_T$ in Eq.~\eqref{eq:integral} to the evaluation of the integrand at the saddle point $b_T^{sp}$. 
When the saddle point is small, $b_T^{sp} \ll  1/\Lambda_{\rm QCD}$, i.e., well in the perturbative region, then one would expect the TMD PDF $F_a(x, k_T^2; Q, Q^2)$ to be mainly controlled by the perturbative physics (always modulo the collinear PDFs). On the contrary, if $b_T^{sp}$ is large, i.e. $b_T^{sp} \gtrsim 1/\Lambda_{\rm QCD}$, the large-$b_T$ non-perturbative contribution is very important and one has to understand/constrain it well, in order to have a full understanding of the TMD PDF. In other words, we use the information on the saddle point $b_T^{sp}$ as an indication of the predictive power of the TMD formalism.

\subsection{Saddle point: general behavior}
To start, we first use the perturbative contribution to $F_a(x, b_T^2; Q, Q^2)$ to compute the saddle-point $b_T^{sp}$. Plugging the perturbative expression in Eq.~\eqref{e:F_pert} into~\eqref{eq:saddle-point}, we obtain
\bea
\frac{d}{db_T} \Bigg\{ & \bigg[\int_{\mu_b}^{Q}\frac{d\mu'}{\mu'}   \left( \ln\left(\frac{Q^2}{\mu'^2}\right) \gamma_K\left(\alpha_s(\mu')\right) - \gamma_F \left( \alpha_s(\mu'), 1 \right) \right)  + D\left(c, \alpha_s(\mu_b)\right) \ln\left( \frac{Q^2}{\mu_b^2}\right) 
\nonumber\\
&- \ln \left(b_T^2 \right)
- \ln \bigg[  \sum_b C_{a/b}(x,b_T^2,\mu_b,\mu_b^2) \otimes f_b(x,\mu_b) \bigg] \Bigg\}\Bigg|_{b_T = b_T^{sp}} = 0\,.
\label{e:sp_equation}
\eea
In general, one can evaluate the saddle point $b_T^{sp}$ of the TMD PDF by solving numerically the above equation. 
This is indeed what we do below when we present the results at next-to-next-to-leading logarithmic (NNLL) accuracy and next-to-next-to-leading order (NNLO) in the strong coupling $\alpha_s$. However, at the leading logarithmic (LL) accuracy where one keeps the leading order (LO) result in $\Gamma_{\rm cusp}$ and in the coefficient functions $C_{a/b}$, one can solve the above equation and obtain the following simple results
\begin{equation}
\label{e:sp_equation_LL}
\frac{1}{2} \ln \left(\frac{Q^2}{\mu_b^{\star\,2}}\right)\ \Gamma_0 \ \frac{\alpha_s(\mu_b^\star)}{4\pi} = 1 - {\cal X}(x,\mu_b^\star) \, , \ \ \ \ \ \ \ \ \ {\cal X}(x,\mu) = \frac{d}{d \ln \mu^2} \ln f_a(x,\mu) \, ,
\end{equation}
where we have introduced $\mu_b^\star = c/b_T^{sp}$.  
The function ${\cal X}(x,\mu)$ quantifies the impact of the DGLAP evolution on the position of the saddle point. Its sign changes according to the value of the light-cone fraction $x$ and determines the $x$-dependence of the saddle point.

The saddle point for the resummed contribution to the Drell-Yan cross section differential with respect to the transverse momentum of the lepton pair has been discussed in Refs.~\cite{Collins:1984kg,Parisi:1979se}. In that treatment the effect of the $x$-dependence was neglected.  
In our treatment, neglecting the $x$-dependence corresponds to setting ${\cal X}=0$. Accordingly, the solution of Eq.~\eqref{e:sp_equation_LL} reads:
\begin{equation}
\label{e:bT_sp_appr_as_0}
b_T^{sp\, (0)} = \frac{c}{\Lambda_{\rm QCD}} \left(  \frac{Q}{\Lambda_{\rm QCD}}  \right)^{-  \Gamma_0 / \left(\Gamma_0 + 8\pi b_0 \right)  } \, , 
\ \ \ \ \ \ \ \ b_0 = \frac{11 C_A - 4 T_f\, n_f }{12 \pi} \, ,
\end{equation}
where $b_0$ is the one-loop coefficient of the QCD beta function~\cite{Patrignani:2016xqp}, and $n_f$ is the number of active flavors. The expression for $b_T^{sp\, (0)}$ is analogous to the one presented in Refs.~\cite{Collins:1984kg,Parisi:1979se}. It follows the usual wisdom that the larger the value of $Q$ is, the smaller $b_T^{sp\, (0)}$ is, and thus the perturbative contributions to the observable play a more important role. 
%
\begin{figure}[h!]
\centering
\begin{tabular}{ccc}
\includegraphics[width=0.475\textwidth]{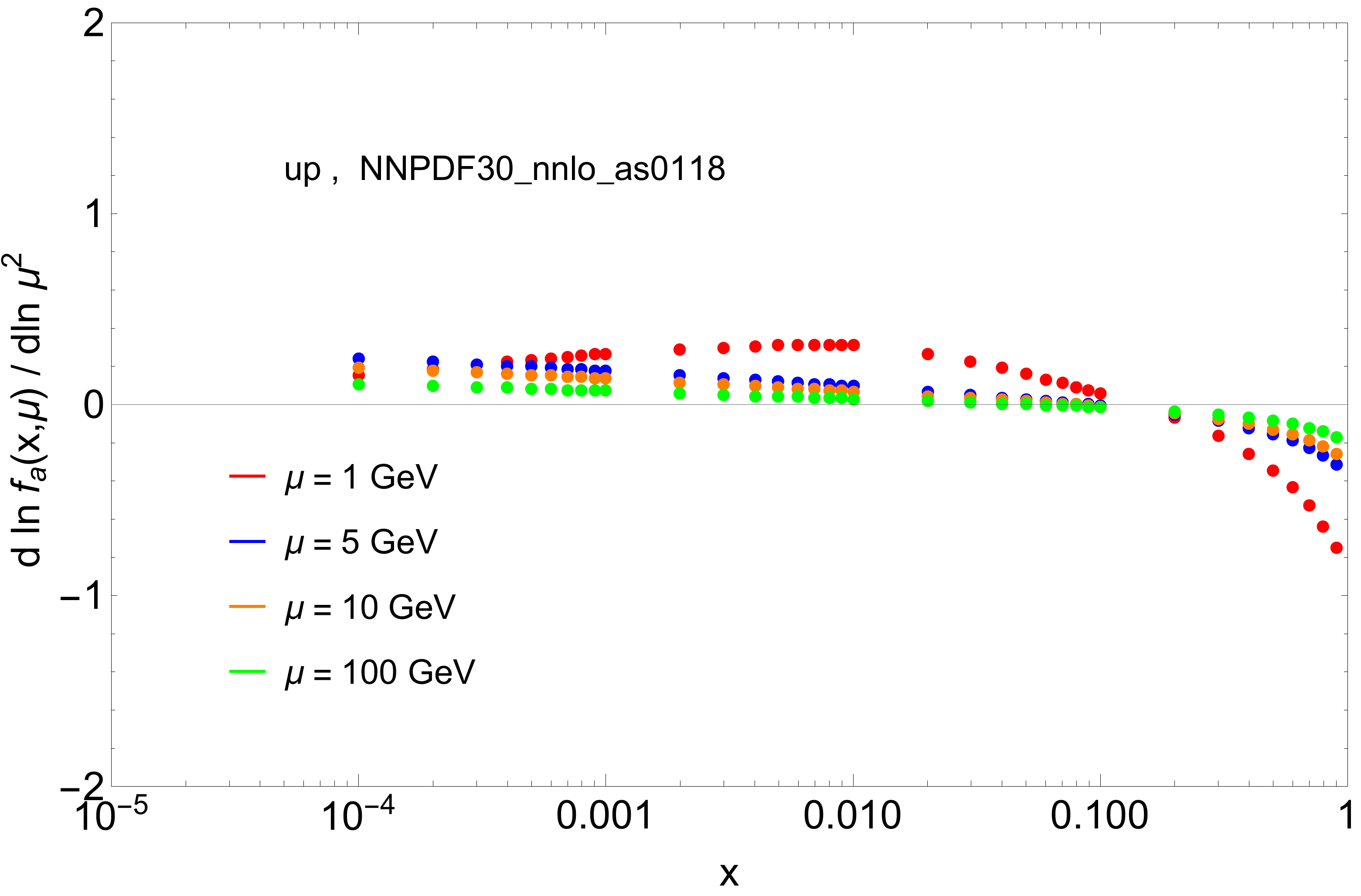} &
\hspace{0.001cm} &
\includegraphics[width=0.475\textwidth]{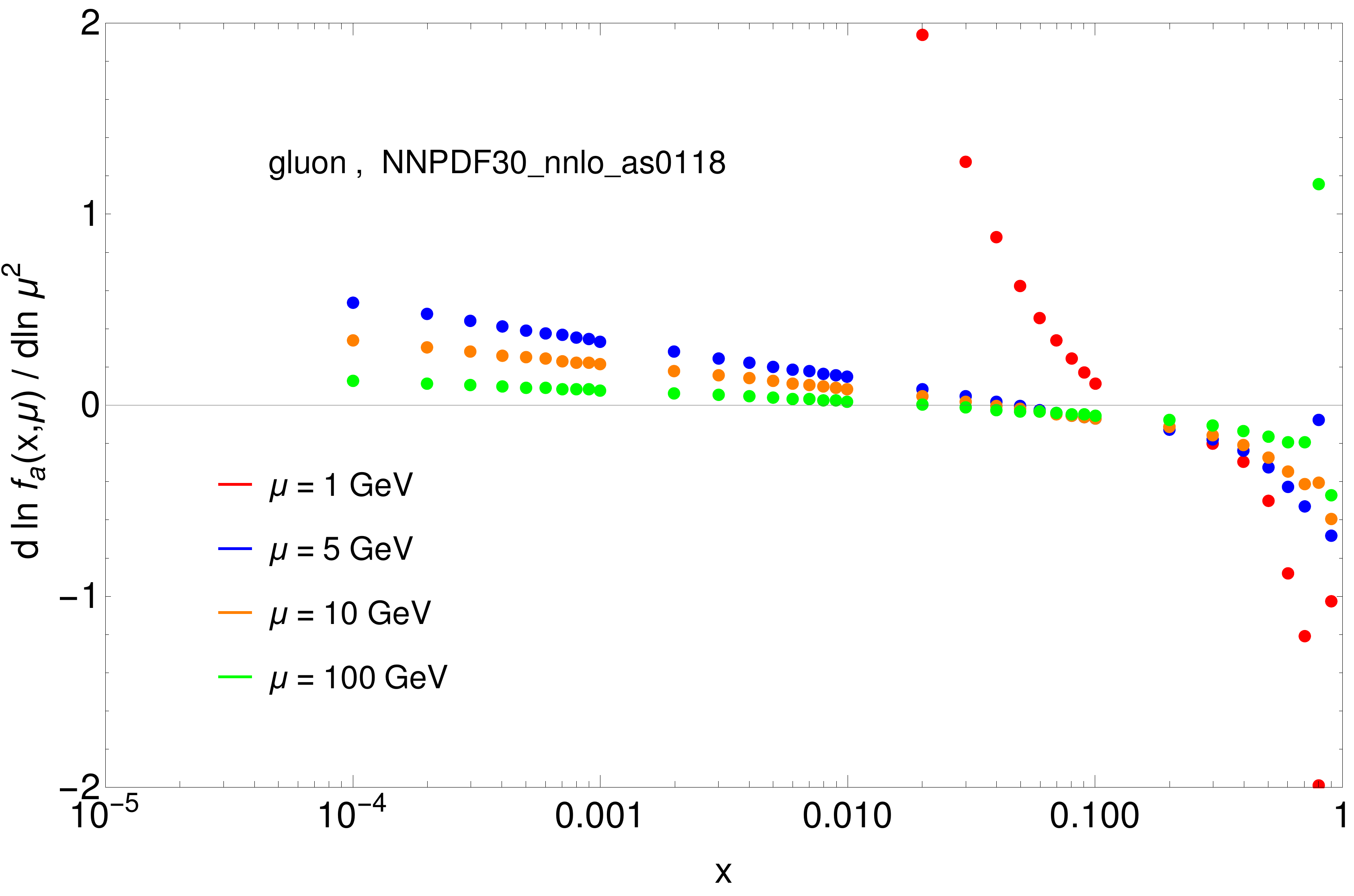} \\
(a) & & (b)
\end{tabular}
\caption{The $x$-dependence of the ${\cal X}(x,\mu)$ function defined in Eq.~\eqref{e:sp_equation_LL} for (a) an up quark, and (b) a gluon. Different values for $\mu=1, 5, 10, 100$~GeV have been chosen.}
\label{f:Xfunct}
\end{figure}
By including the contribution of ${\cal X}$, the solution to Eq.~\eqref{e:sp_equation_LL} acquires an $x$-dependence: 
\begin{equation}
\label{e:bT_sp_appr_as}
b_T^{sp} = \frac{c}{\Lambda_{\rm QCD}} \left(  \frac{Q}{\Lambda_{\rm QCD}}  \right)^{-  \Gamma_0 / \left[\Gamma_0 + 8\pi b_0 \left(1 - {\cal X}(x,\mu_b^\star) \right)  \right]  } \ . 
\end{equation}
Note that the right-hand side of Eq.~\eqref{e:bT_sp_appr_as} depends on $b_T^{sp}$ through $\mu_b^\star$, and thus Eq.~\eqref{e:sp_equation_LL} needs to be solved by iterations. A legitimate choice for the first iteration is to evaluate ${\cal X}$ at $b_T^{sp\, (0)}$. Comparing Eq.~\eqref{e:bT_sp_appr_as} with \eqref{e:bT_sp_appr_as_0}, one observes that if ${\cal X} > 0$ ($<0$), one would have $b_T^{sp} < b_T^{sp\,(0)}$ ($b_T^{sp} > b_T^{sp\,(0)}$). To understand the behavior of ${\cal X}$, as well as for the general numerical investigation, below we rely on the {\tt LHAPDF6} library~\cite{Buckley:2014ana} and in particular on the central PDF set from {\tt NNPDF30}~\cite{Ball:2014uwa} at NNLO accuracy with $\alpha_s(M_Z) = 0.118$. We also use the {\tt APFEL} library~\cite{Bertone:2013vaa} to calculate the ${\cal X}$ function. The result is in agreement with applying the finite differences method to the {\tt NNPDF30} grid. In Fig.~\ref{f:Xfunct}, we plot ${\cal X}$ as a function of $x$ for an up quark (left) and a gluon (right), at different scales $\mu = 1, 5, 10, 100$~GeV, respectively. Apart from the gluon case at $\mu\lesssim 1$~GeV, the function ${\cal X}$ is positive for $x \lesssim 0.1$ and negative for $x \gtrsim 0.1$. 
Thus its effect is to reduce the value of the saddle point $b_T^{sp}$ with respect to the solution $b_T^{sp\, (0)}$ for $x \lesssim 0.1$ and to increase it for $x \gtrsim 0.1$. Because of this, for the same $Q$ value but smaller $x$ region, the perturbative contribution (from small-$b_T$ region) plays a more important role for the TMD PDF. This means that in general, away from the limiting cases, the TMD PDF is more perturbatively dominated at large $Q$ and small $x$. On the other hand, the TMD PDF is more dominated by the non-perturbative contribution at small $Q$ and large $x$. This suggests that even for a moderately large $Q$, the TMD PDF at large $x$ could become quite sensitive to the non-perturbative contribution, due to the $x$-dependence of the ${\cal X}$ function. 

\subsection{Saddle point: detailed analysis}
\label{ss:behavior_sol}
After the above qualitative understanding of the kinematic dependence of the saddle point, we now turn to a detailed numerical analysis and concentrate on the $x$ and $Q$ dependence. We first choose representative values of $x$, and study the $Q$-dependence of the saddle point $b_T^{sp}$. For the small $x$ region, we choose $x=10^{-3}$ which could be relevant to the LHC and the EIC kinematics. While for large $x$ region, we choose $x=0.3$ for our illustration below. 

\begin{figure}[h!]
\centering
\begin{tabular}{ccc}
\includegraphics[width=0.475\textwidth]{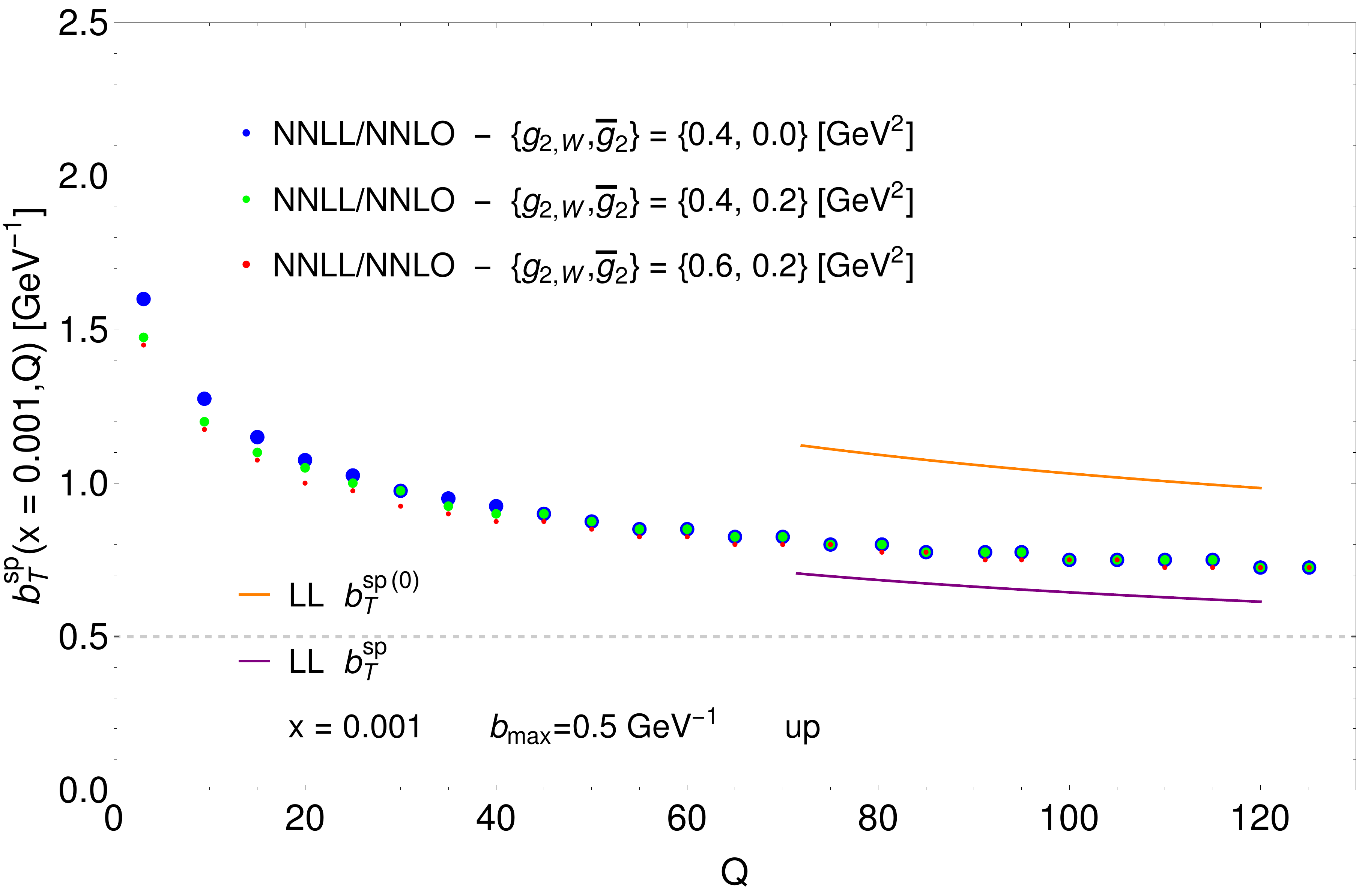} &
\hspace{0.001cm} &
\includegraphics[width=0.475\textwidth]{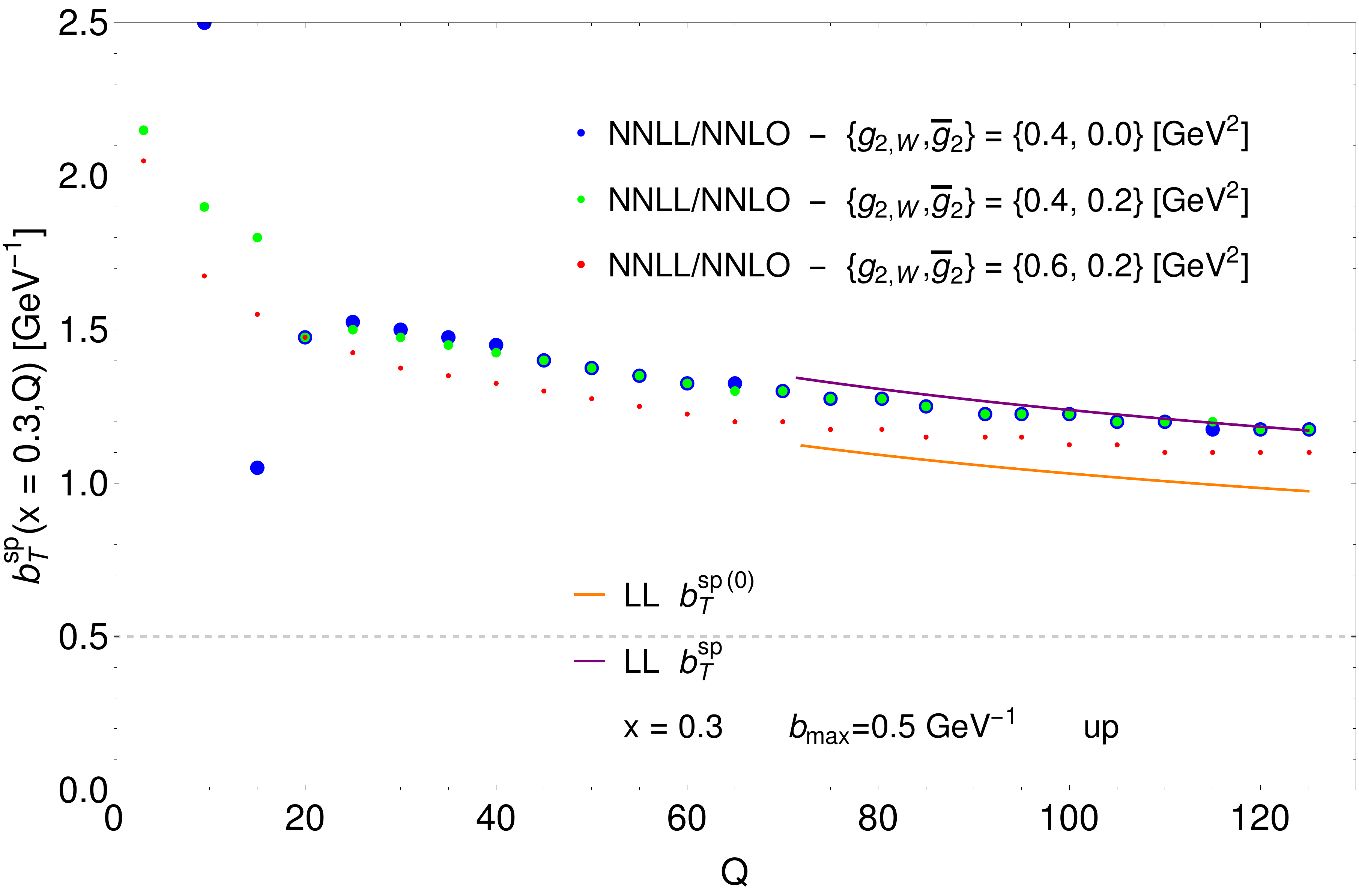} \\
(a) & & (b) \\ \\ \\ 
\includegraphics[width=0.475\textwidth]{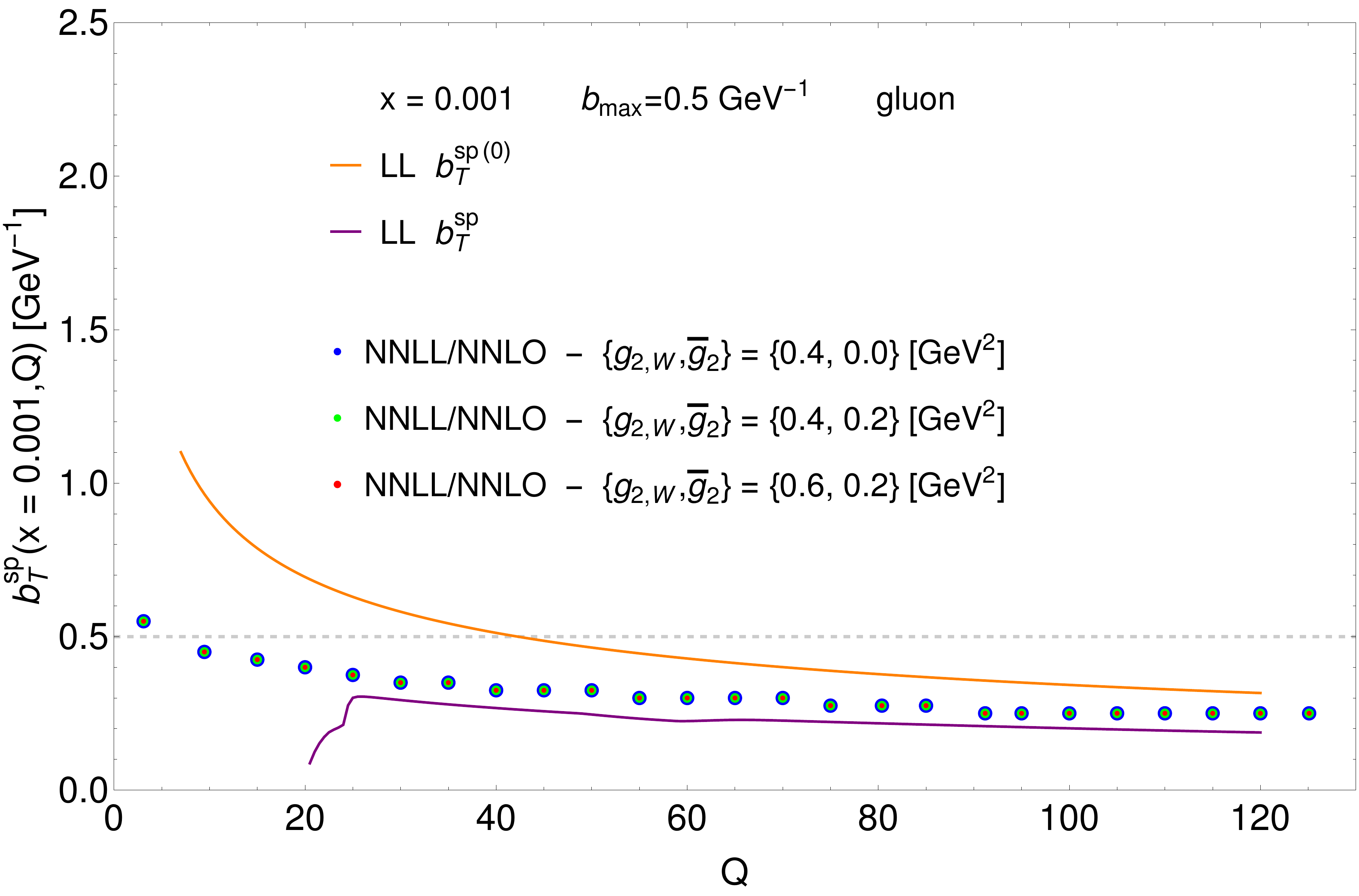} &
\hspace{0.001cm} &
\includegraphics[width=0.475\textwidth]{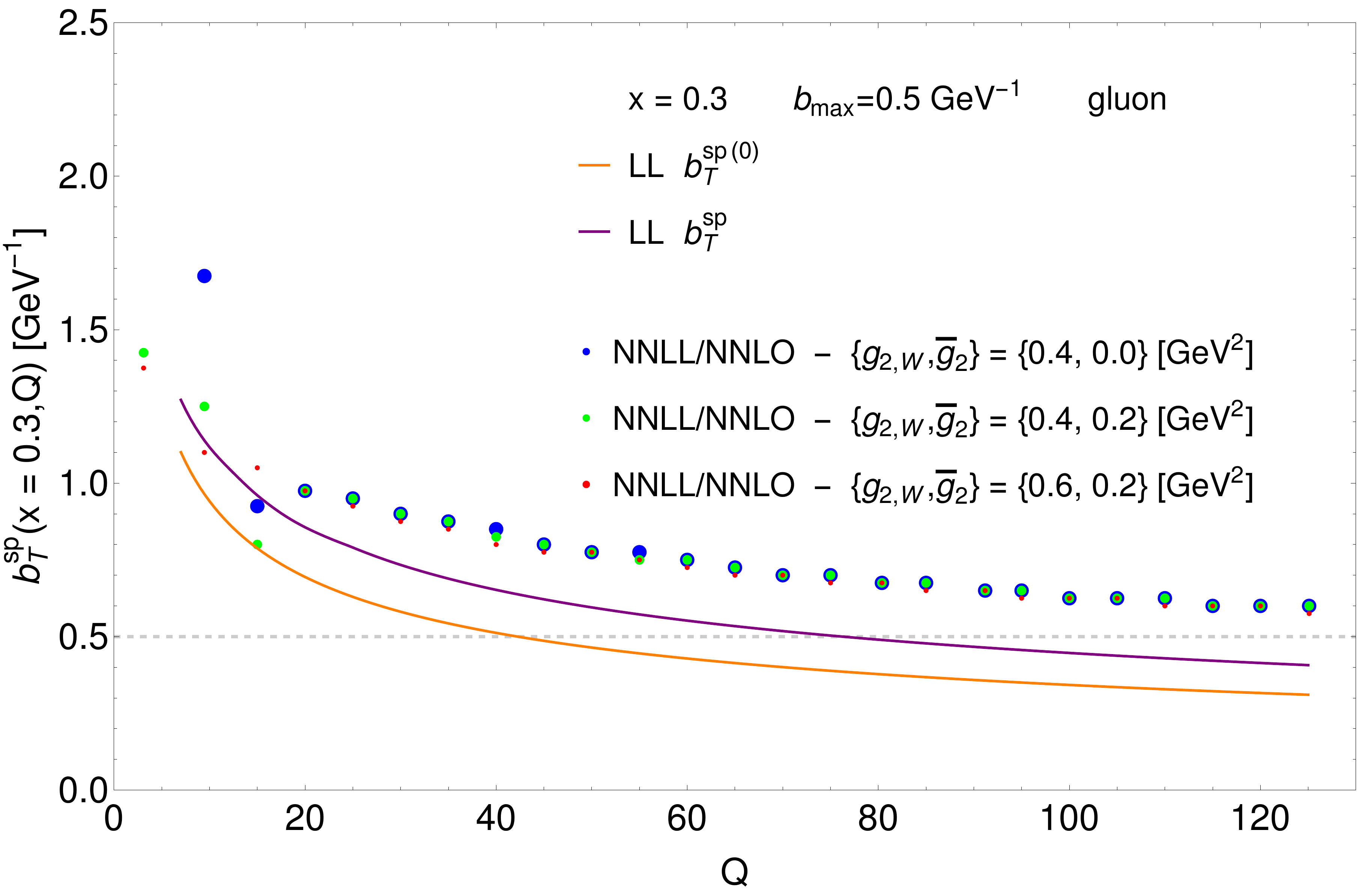} \\
(c) & & (d)
\end{tabular}
\caption{Position of the saddle point for the TMD PDF as a function of the scale $Q$ for: (a) up quark at $x = 10^{-3}$, (b) up quark at $x = 0.3$, (c) gluon at $x = 10^{-3}$, and (d) gluon at $x = 0.3$. The orange and the purple curves represent the analytic leading log solutions, without and with $x$-dependence respectively. The dots corresponds to the numerical studies of the saddle point including higher-orders and the large $b_T$ corrections.
The behavior of the $b_T^{sp}$ solution for the gluon at low $x$ presents a non-smooth behavior towards the low $Q$ region due to the non-smooth behavior of the corresponding ${\cal X}$ function.}
\label{f:Qdep_sp}
\end{figure}

Let us first plot the behavior of the $x$-independent solution $b_T^{sp\, (0)}$ and the $x$-dependent one, $b_T^{sp}$, both at LL accuracy as given in Eqs.~\eqref{e:bT_sp_appr_as_0} and \eqref{e:bT_sp_appr_as}. In Fig.~\ref{f:Qdep_sp} the orange curves represent $b_T^{sp\, (0)}$, whereas the purple curves refer to $b_T^{sp}$ as a function of the hard scale $Q$: (a) up quark at small $x = 10^{-3}$, (b) up quark at large $x = 0.3$, (c) gluon at small $x = 10^{-3}$, and (d) gluon at large $x = 0.3$. Note that when $b_T^{sp\, (0)} > c = 2e^{-\gamma_E}$, the first iteration in the solution of Eq.~\eqref{e:bT_sp_appr_as} is evaluated at the scale $\mu_b^\star(b_T^{sp\, (0)}) < 1$ GeV. Thus, the collinear PDF $f_a(x,\mu)$ is evaluated, by extrapolation, at a scale below 1 GeV, where the used  phenomenological parametrization is not to be trusted. The same applies to any other iteration to calculate $b_T^{sp}$. For this reason, the orange and purple curves are displayed only when $b_T^{sp\, (0)} < 2e^{-\gamma_E}$. 

%

Several comments are in order. First of all, both orange and purple curves are decreasing as $Q$ increases, as expected. Just as we have emphasized in the previous section, as $Q$ increases, the saddle point, both for $b_T^{sp\, (0)}$ and $b_T^{sp}$, becomes smaller indicating that the perturbative contribution becomes more important. Second of all, one can see clearly for the small-$x$ region that the purple curves are below the orange curves, i.e., $b^{sp}_T< b^{sp\,(0)}_T$. This is driven by the contribution of a positive ${\cal X}$ as discussed in the previous section. Similarly, for the large-$x$ region, the purple curves are above the orange curves, i.e., $b^{sp}_T> b^{sp\,(0)}_T$, again consistent with our analysis above. 

\begin{figure}[tbh]
\centering
\begin{tabular}{ccc}
\includegraphics[width=0.475\textwidth]{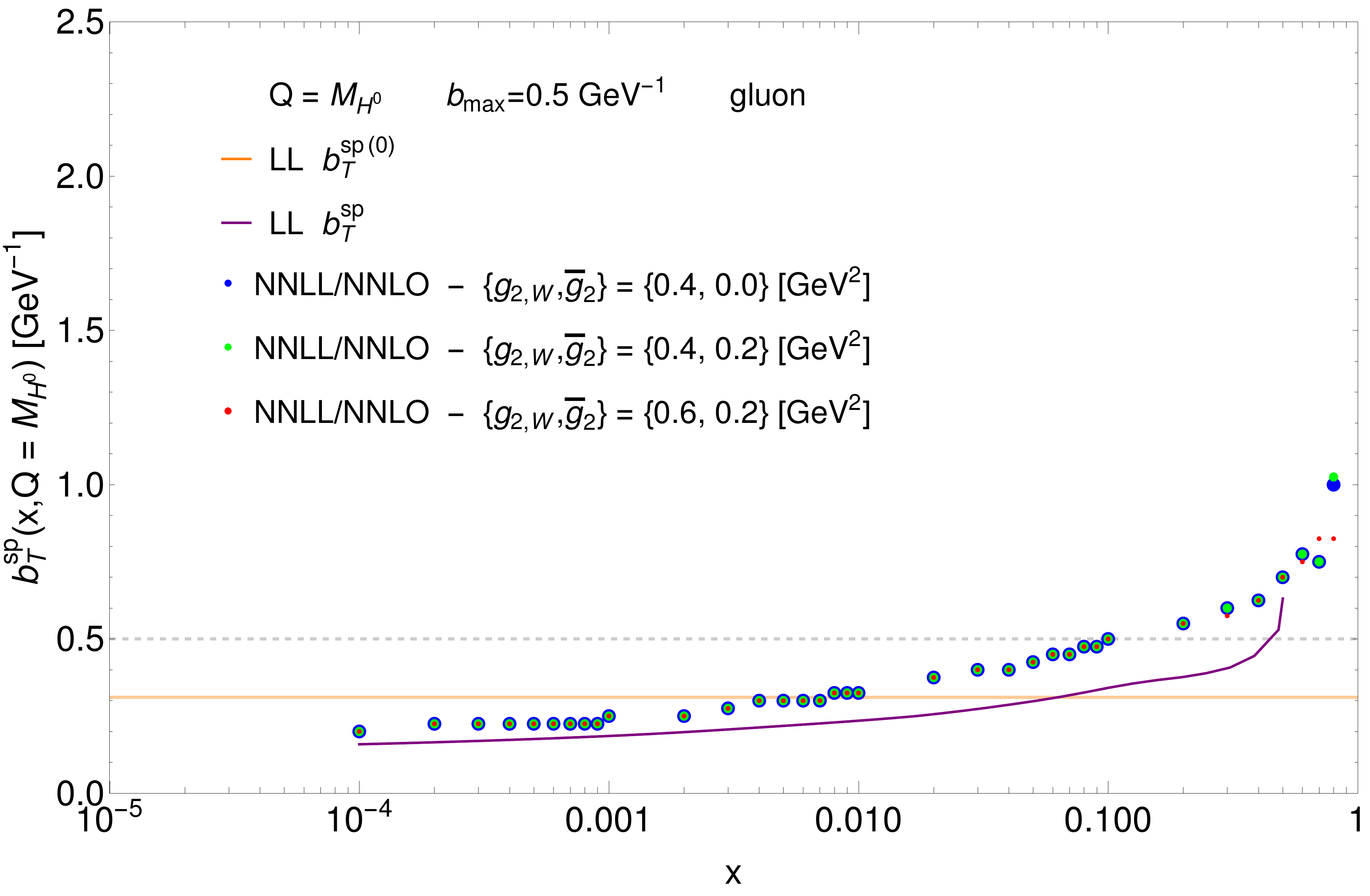} &
\hspace{0.001cm} &
\includegraphics[width=0.475\textwidth]{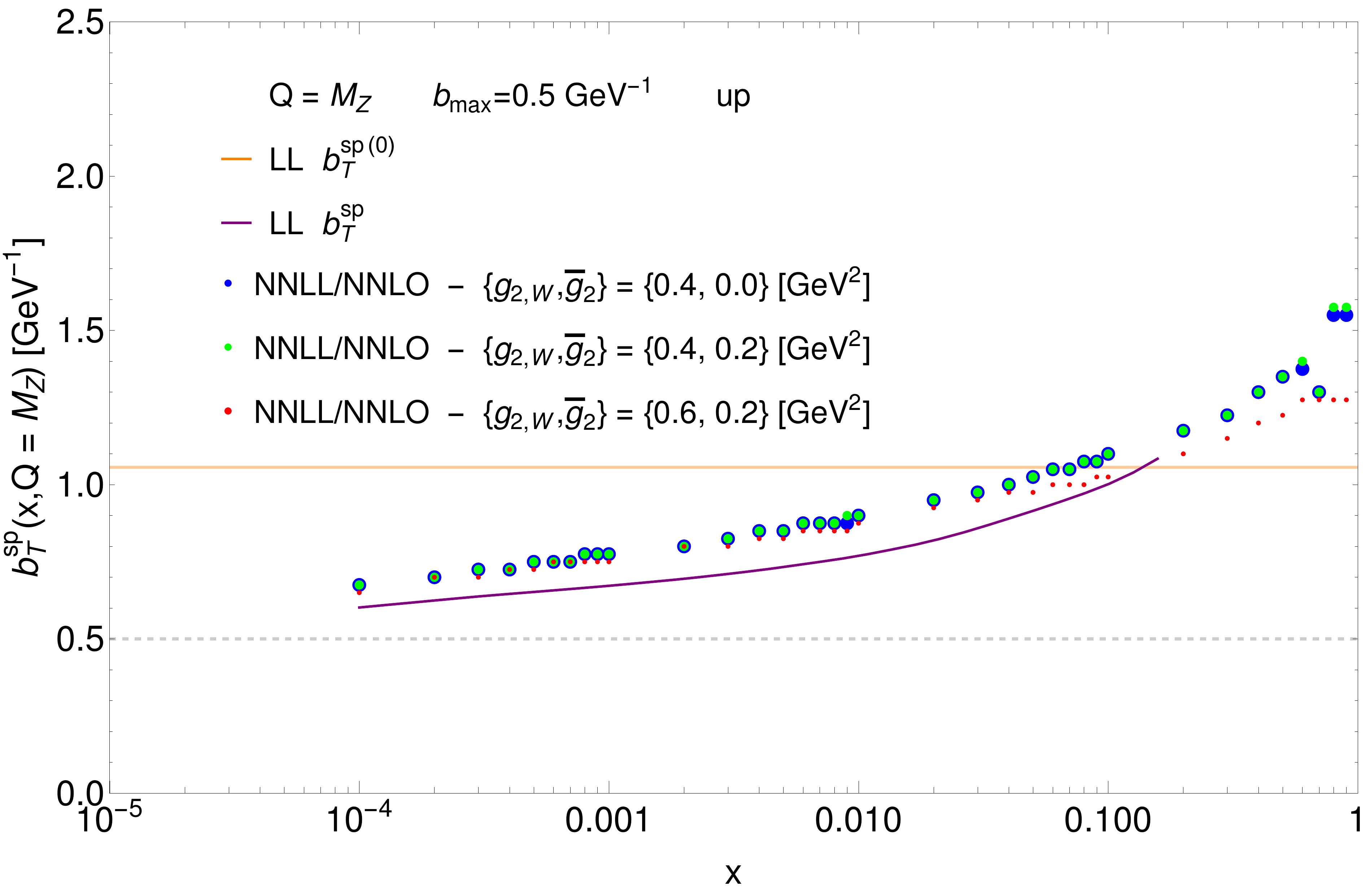} \\
(a) & & (b) \\  \\  
\includegraphics[width=0.475\textwidth]{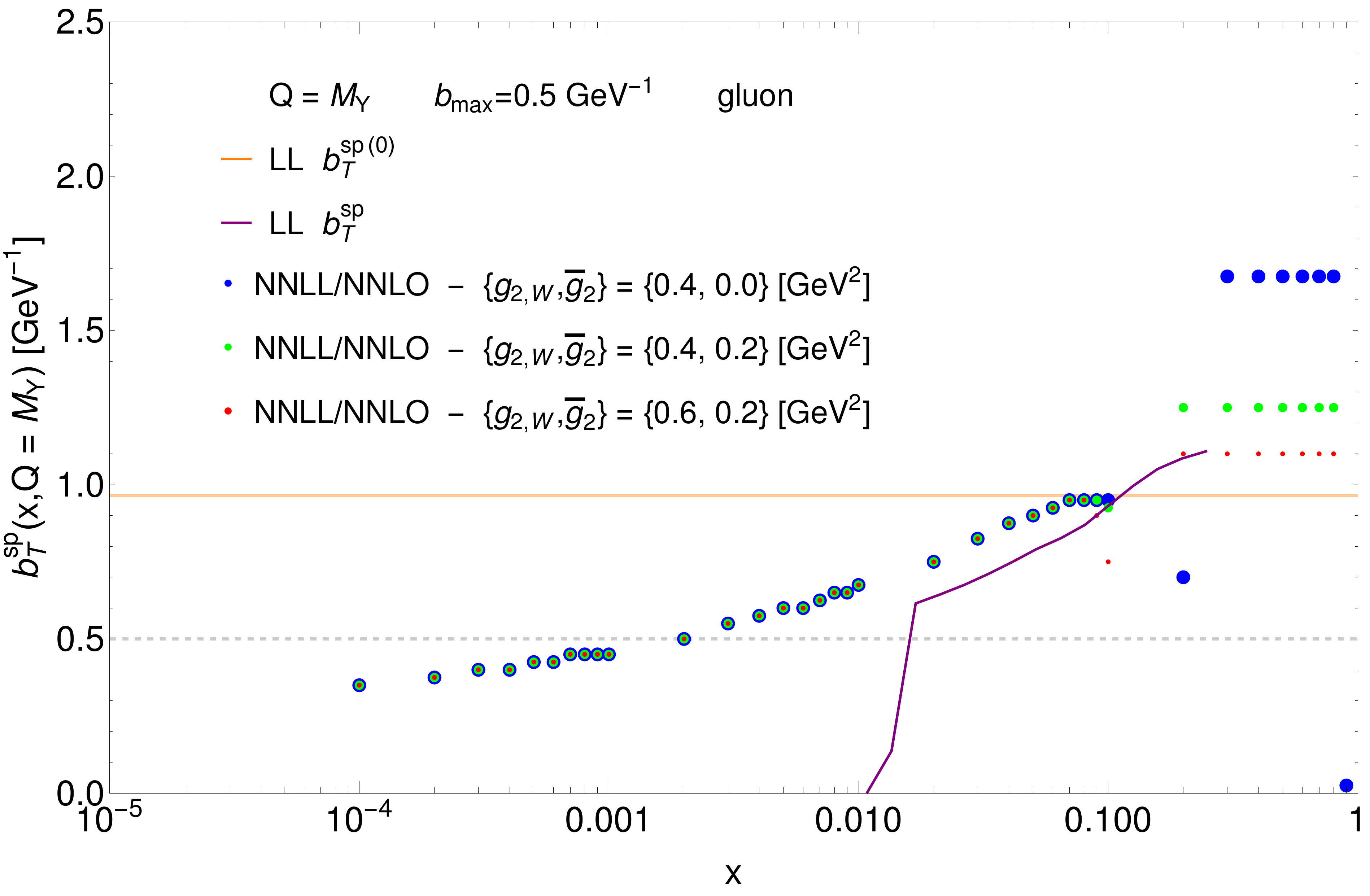} &
\hspace{0.001cm} &
\includegraphics[width=0.475\textwidth]{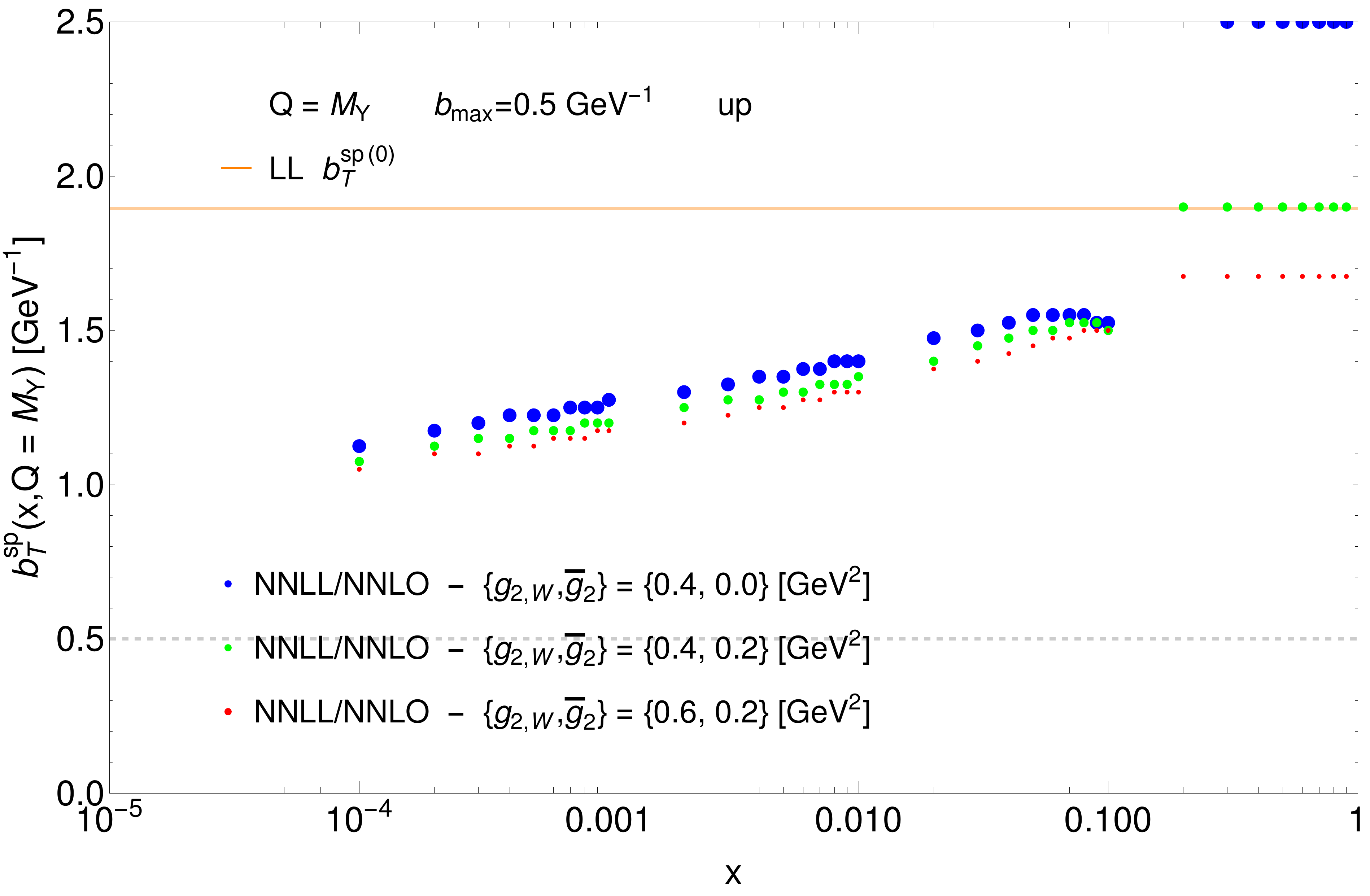} \\
(c) & & (d) \\ \\ 
\includegraphics[width=0.475\textwidth]{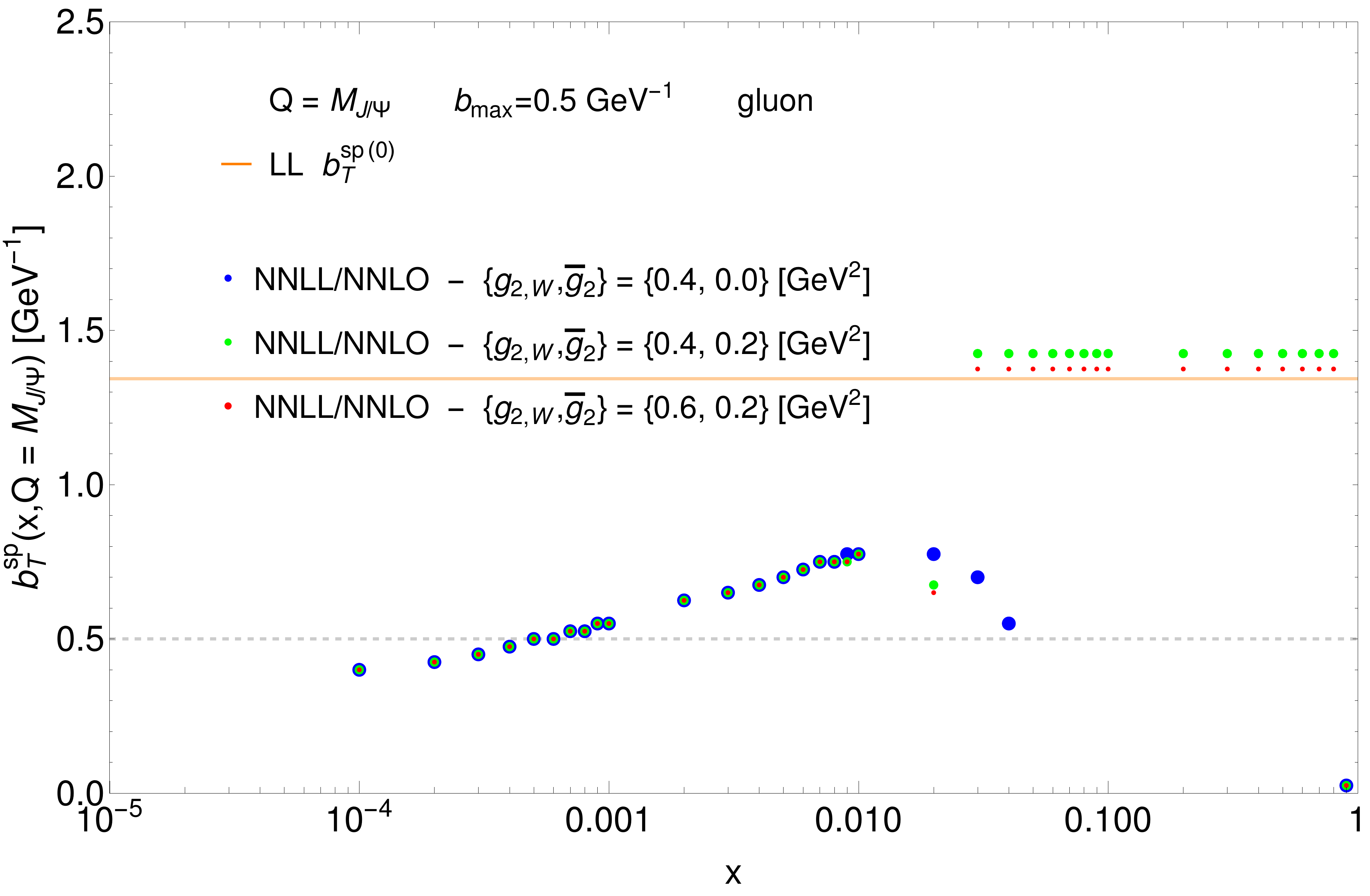} &
\hspace{0.001cm} &
\includegraphics[width=0.475\textwidth]{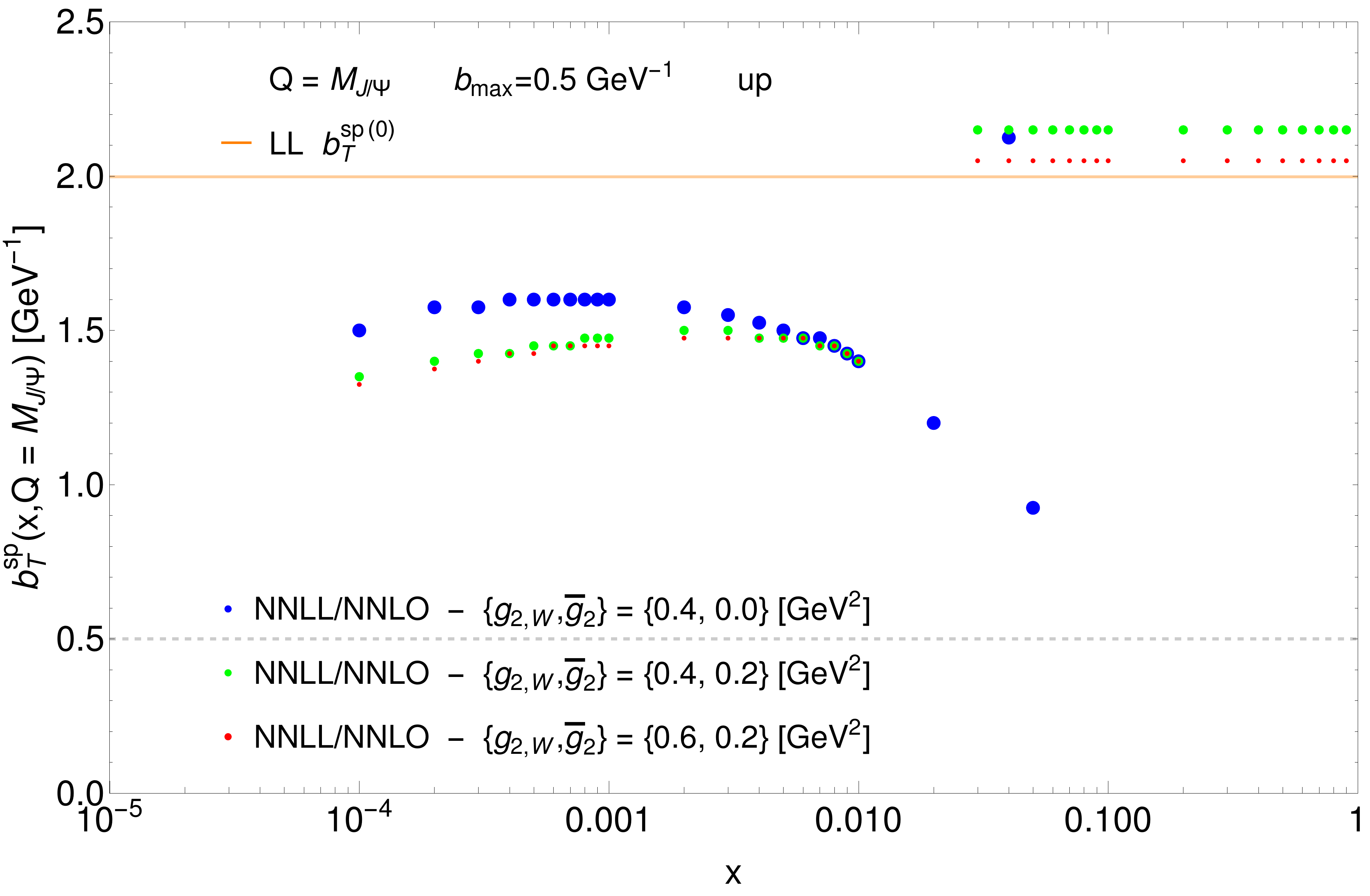} \\
(e) & & (f)
\end{tabular}
\caption{
Position of the saddle point for a TMD PDF as a function of the light cone momentum fraction $x$ for: 
(a) gluon at $Q = M_{H^0} = 125$ GeV,  
(b) up quark at $Q = M_{Z} = 91$ GeV,  
(c) gluon at $Q = M_{\Upsilon} = 9.46$ GeV, 
(d) up quark at $Q = M_{\Upsilon}$, 
(e) gluon at $Q = M_{J/\Psi} = 3.096$ GeV, and 
(f) up quark at $Q = M_{J/\Psi}$.
The behavior of the saddle point from high energies (top) to low energies (bottom) is summarized, for the gluon case (left) and the up quark case (right). The characteristics are discussed in detail in Sec.~\ref{s:saddle}. 
The overall trend is that the saddle point for a gluon lies at lower $b_T$ values with respect to the quark case at equal or comparable energy scales. 
The $x$ dependence induced by the perturbative structure of the TMD PDF is monotonically increasing, and deviations from this trend are generated by the treatment of the large $b_T$ region.}
\label{f:xdep_sp}
\end{figure}

The parameter $\bmax = 0.5$ GeV$^{-1}$ in principle identifies the perturbative region $b_T < \bmax$, but, considering that this is an arbitrary choice, we can allow for some degree of tolerance and identify the ``extended'' perturbative region as $b_T < 1$ GeV$^{-1}$. 
In terms of detailed numerical values, we find from Fig.~\ref{f:Qdep_sp} that for the small-$x$ region, the purple curve for up quark is below 1~GeV$^{-1}$, i.e. $b_T^{sp} < 1$~GeV$^{-1}$ when $Q\gtrsim 60$ GeV, indicating that the perturbative or small-$b_T$ contribution plays a more important role for the up quark TMD PDF in the small-$x$ region. On the other hand, for the large-$x$ region, even when $Q\gtrsim 120$ GeV, the saddle point is still larger than 1~GeV$^{-1}$, suggesting that the non-perturbative or large-$b_T$ contribution would still play a significant role for the up quark TMD PDF in the large-$x$ region, even though the $Q$ value is already very large. Similar observations apply to the gluon TMD PDF, in an even better way. Due to the larger color factor ($C_A$ vs $C_F$) in $\Gamma_{\rm cusp}$, the Sudakov factor makes the gluon TMD PDF more narrowly concentrate in the small-$b_T$ region. For example, for a gluon TMD PDF in both the small and large-$x$ regions, the saddle point $b_T^{sp}$ would become smaller than 1~GeV$^{-1}$ for moderate $Q\gtrsim 20$ GeV already, suggesting that the non-perturbative contribution plays a less important role in determining the gluon TMD PDF. 
We also note that the $x$-dependent LL solution for the gluon at low $x$ becomes non-smooth in the low $Q$ region (see Fig.~\ref{f:Qdep_sp}~(c) and Fig.~\ref{f:xdep_sp}~(c)): this is essentially due to the non-smooth behavior of the ${\cal X}$ function. 

Let us analyze the saddle point by including the extrapolation term $R_a^{\rm NP}(x, b_T, Q; b_{\rm max})$ in Eqs.~\eqref{e:piecewise_TMD} and~\eqref{eq:R-np}. For that, we evaluate the saddle point of the TMD PDF by directly solving numerically Eq.~\eqref{eq:saddle-point} at NNLL and NNLO. In such a setup, we include $\Gamma_{0,1,2}$ and $\gamma_{0,1}$ in the anomalous dimension, and use two-loop results for the coefficient functions $C_{a/b}$, as given in Ref.~\cite{Echevarria:2016scs}. As previously discussed, there are four parameters in the extrapolation function $R_a^{\rm NP}(x, b_T, Q; b_{\rm max})$, namely $\alpha,~g_1,~g_2$, and ${\bar g}_2$. In Ref.~\cite{Qiu:2000hf}, the following quantity is defined 
\bea
\label{e:g2p_def}
g_{2}^\prime(Q) \equiv {\bar g}_2 + g_2 \ln\left(\frac{Q^2}{\mu_{\bmax}^2}\right)\,,
\eea
and its value at the scale of the $W$ boson mass  
is determined to be $g_{2,W} = 0.4$ GeV$^2$ through a fit to the experimental data~\cite{Qiu:2000hf}. From a given value of $g_{2,W}$ and $\bar{g}_2$, the value of $g_2$ can be determined by inverting the above equation. In the analysis below, we either fix the value of $g_2$ or $g_{2,W}$, or vary around their value. As we have mentioned before, we require $F_a(x,b_T^2; Q, Q^2)$ to be smooth at $b_T=\bmax$ so to determine the other parameters in the extrapolation function $R_a^{\rm NP}$. Specifically, we require the first and second order derivatives of $F_a(x,b_T^2; Q, Q^2)$ to be continuous at $b_T=\bmax$. With two conditions, two of the parameters can be fixed and we choose to be $\alpha$ and $g_1$.  
There is a subtlety here that requires some caution. In the context of this analysis, which is focused on the high energy regime, we determine $\alpha$ and $g_1$ through the continuity of the first and second derivative only if the first derivative in $b_T=\bmax$ is negative ($\partial F_a(x,b_T^2=\bmax^2;Q,Q^2)/ \partial b_T < 0$) and thus $F_a(x,b_T^2;Q,Q^2)$ decreases as $b_T$ increases to be consistent with the expected physical behavior. 
On the contrary, which is usually the case at very large $x$, when such a first derivative is positive, we set $\alpha$ and $g_1$ to zero. This is one of the possible methods to avoid an unphysical extrapolation in the large $b_T$ region. Other more flexible strategies that can guarantee non-zero values for $\alpha$ and $g_1$ can be introduced in order to describe events at low $Q$, for example in the context of Semi-Inclusive Deep-Inelastic Scattering at fixed-target energies. We leave such a detailed analysis for future studies. 

In Fig.~\ref{f:Qdep_sp}, we plot the saddle point $b_T^{sp}$ for three different scenarios: (1) $(g_{2,W}, {\bar g}_2) = (0.4, 0.0)$~GeV$^2$, denoted as blue dots, (2) $(g_{2,W}, {\bar g}_2) = (0.4, 0.2)$~GeV$^2$, denoted as green dots, (3) $(g_{2,W}, {\bar g}_2) = (0.6, 0.2)$~GeV$^2$, denoted as red dots. It is evident for the small-$x$ and large-$Q$ region that the numerical values of the saddle points are quite stable for both quarks and gluons, such as at $Q=M_Z=91.18$ GeV ($Z$ boson) and $Q=M_{H^0}=125.1$ GeV (Higgs boson). This suggests that the non-perturbative contributions are mild in these cases. On the other hand, for the quark TMD PDF in the large-$x$ region, the red dots can be different from the blue/green dots even for very large-$Q$ values, suggesting that the non-perturbative contribution could be quite significant. On the other hand, the situation is quite improved for the gluon TMD PDF at large-$x$, thanks to the strong Sudakov resummation effect. A certain degree of model dependence is left for the gluon at large $x$ and small $Q$, which anyway vanishes for the gluon at small $x$, where the saddle point is almost exclusively in the strict perturbative region $b_T < \bmax$.

In Fig.~\ref{f:xdep_sp}, we plot the position of the saddle point for a TMD PDF as a function of the light cone momentum fraction $x$ for: 
(a) gluon at $Q = M_{H^0}$,  
(b) up quark at $Q = M_{Z}$,  
(c) gluon at $Q = M_{\Upsilon} = 9.46$ GeV, 
(d) up quark at $Q = M_{\Upsilon}$, 
(e) gluon at $Q = M_{J/\Psi} = 3.096$ GeV, 
(f) up quark at $Q = M_{J/\Psi}$.
The behavior of the saddle point from high energies (top) to low energies (bottom) is summarized, for the gluon case (left) and the up quark case (right). 
At this point it is important to remark that the $x$-dependence of the numerical solutions (the dots) for $b_T^{sp}$ in Fig.~\ref{f:xdep_sp} is driven both by the $x$-dependence of the perturbative part and of the non-perturbative part ($R_a^{NP}$) of the TMD PDF via $g_1$ and $\alpha$. Indeed, when $b_T > \bmax$, if one sets manually $g_1$ and $\alpha$ to zero, the $x$ dependence is lost. 
As previously discussed, this is also what happens at (very) large $x$ in all cases apart for the gluon at $Q=M_{H^0}$, when the first derivative of the TMD PDF at $b_T=\bmax$ becomes positive.  
The $x$ dependence generated by the perturbative contribution is generally monotonically increasing. A confirmation of this trend can be found in the shape of the ${\cal X}$ function. Thus, the changes in concavity in the large $x$ regions are essentially induced by the treatment of the large $b_T$ region and thus model dependent. 

The overall trend that we can infer from Fig.~\ref{f:xdep_sp} is that the saddle point for a gluon lies at lower $b_T$ values with respect to the quark case at equal or comparable energy scales, again due to the different color factor in the cusp anomalous dimension. 
It is instructive to point out that, for physical observables which depend on the convolution in momentum space of two TMD PDFs, such as the transverse momentum differential cross section of $W/Z$ and $H^0$ boson production, the integrand in the $b_T$-space is more peaked in the low $b_T$ region than for the single TMD PDF. Thus at large $Q$ and small $x$ region, the predictive power is then guaranteed (see Fig.~\ref{f:sigma_Z} in Sec.~\ref{s:cross_sections} and Refs.~\cite{Qiu:2000hf,Berger:2002ut,Berger:2003pd,Berger:2004cc}). 

In practice, the plots in Fig.~\ref{f:xdep_sp} suggest that the transverse momentum distribution of $H^0$ and $Z$ bosons at small-$x$ (or large center-of-mass energy $\sqrt{s}$) would be very well controlled by the perturbative contribution. 
If we are in the small-$x$ region while at the moderate scale of $\Upsilon$ mass, $M_{\Upsilon}$, the non-perturbative contribution to the gluon TMD PDF could be mild. This suggests that the transverse momentum distribution of the $\Upsilon$ particle could be very well described by the perturbative physics at the collider energy such as the LHC ~\cite{Berger:2004cc,Qiu:2017xbx}, where the gluon-gluon fusion channel dominates the production cross section, but not at lower energies. 
Finally, for the $J/\psi$ production, which is at a very low mass scale $M_{J/\psi}$ GeV, the non-perturbative contribution would be more important and could be even entangled with the formation of the quarkonium~\cite{Echevarria:2019ynx,Fleming:2019pzj}. 
For the quark case, the predictive power is well under control at $Q=M_Z$, as we shall see in Sec.~\ref{s:cross_sections}, whereas the physical observables receive significant non-perturbative corrections for $Q \lesssim 10$ GeV. 

Overall we can conclude that the kinematic domain in which the predictive power is strongest is the large-$Q$ and small-$x$ region, where the saddle point $b_T^{sp}$ for the transverse momentum distribution is comparable to or smaller than 0.5 GeV$^{-1}$. We emphasize again that in addition to the value of the hard scale $Q$, this analysis shows that also the value of the light-cone fraction $x$ contributes to determining how relevant the non-perturbative part of the TMD PDF is. 
This is essential also to understand which experiments and kinematic configurations can be more useful to investigate the properties of the non-perturbative structure of hadrons and which other experimental configurations are more suited for testing the predictive power of the theory. 
It is certainly important to keep in mind that the predictive power of any theory always depends also on the precision of the specific observable studied in order to test and falsify the theory itself (see Sec.~\ref{ss:Z_prod}).

\section{Relevance of non-perturbative corrections}
\label{s:NP_relevance}
Apart from the saddle point of the TMD PDF, 
it is also useful to directly look at the integrand in $b_T$-space of the TMD PDF at $k_T=0$ , which is simply 
\bea
b_T\, F_a(x,b_T^2; Q, Q^2)/2\pi\, .
\eea
The shape of this function is also useful to quantify the relevance of the large-$b_T$ part of the TMD PDF. In this section, we will assess the relevance of non-perturbative contributions more quantitatively.

\begin{figure}[h!]
\centering
\begin{tabular}{ccc}
\includegraphics[width=0.475\textwidth]{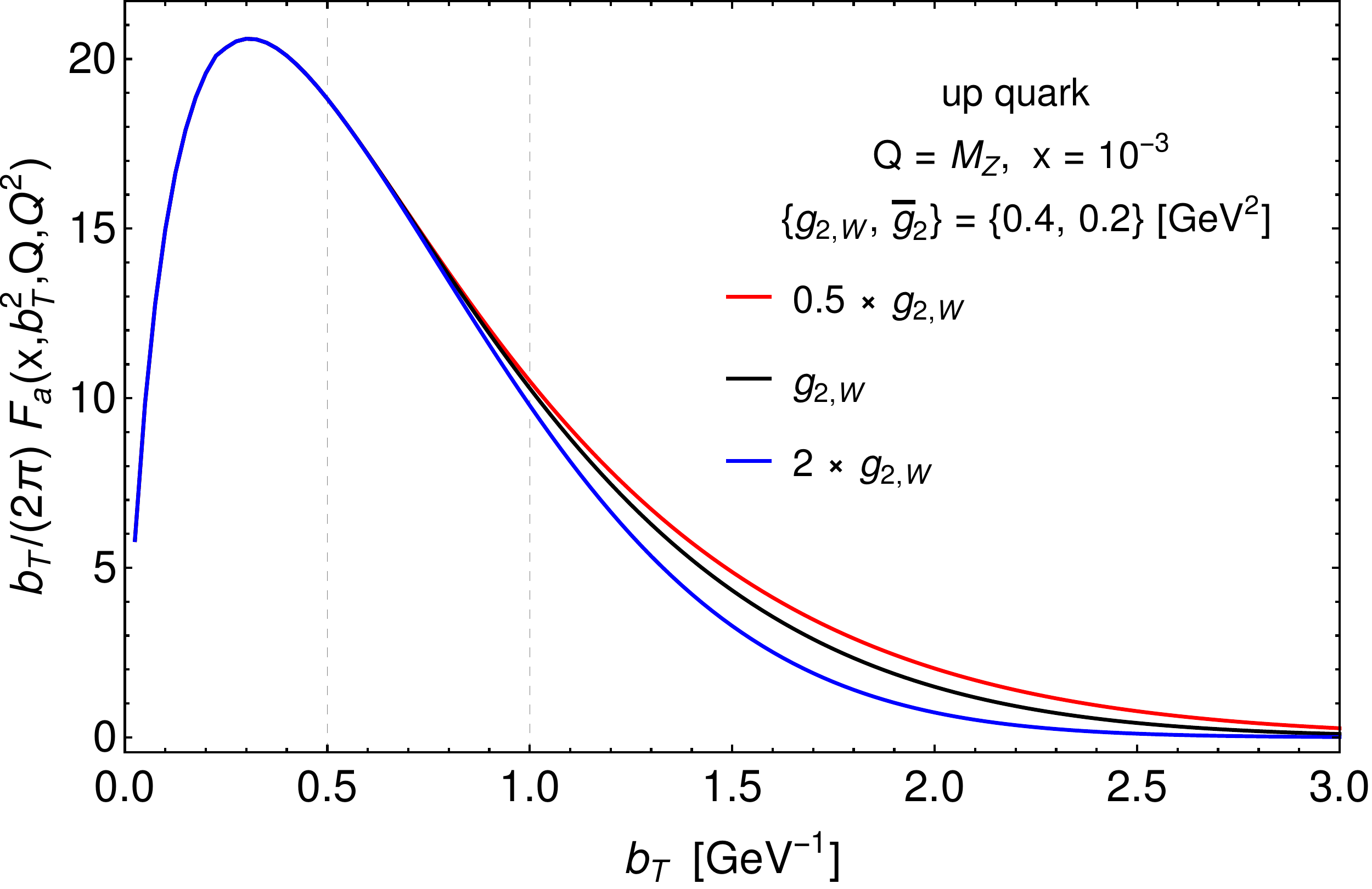}
&\hspace{0.001cm}
&
\includegraphics[width=0.475\textwidth]{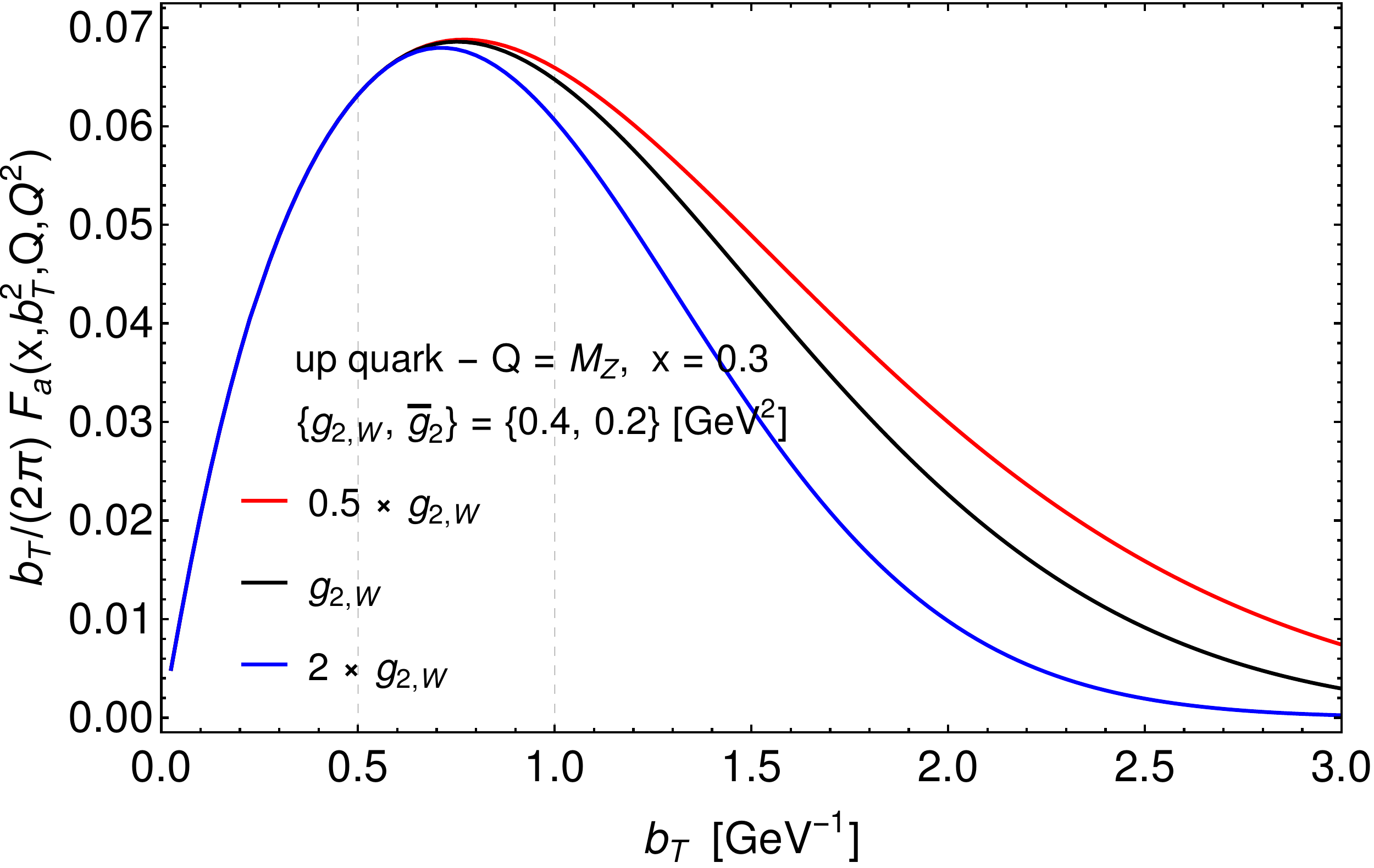}
\\
(a) && (b)
\\ \\
\includegraphics[width=0.475\textwidth]{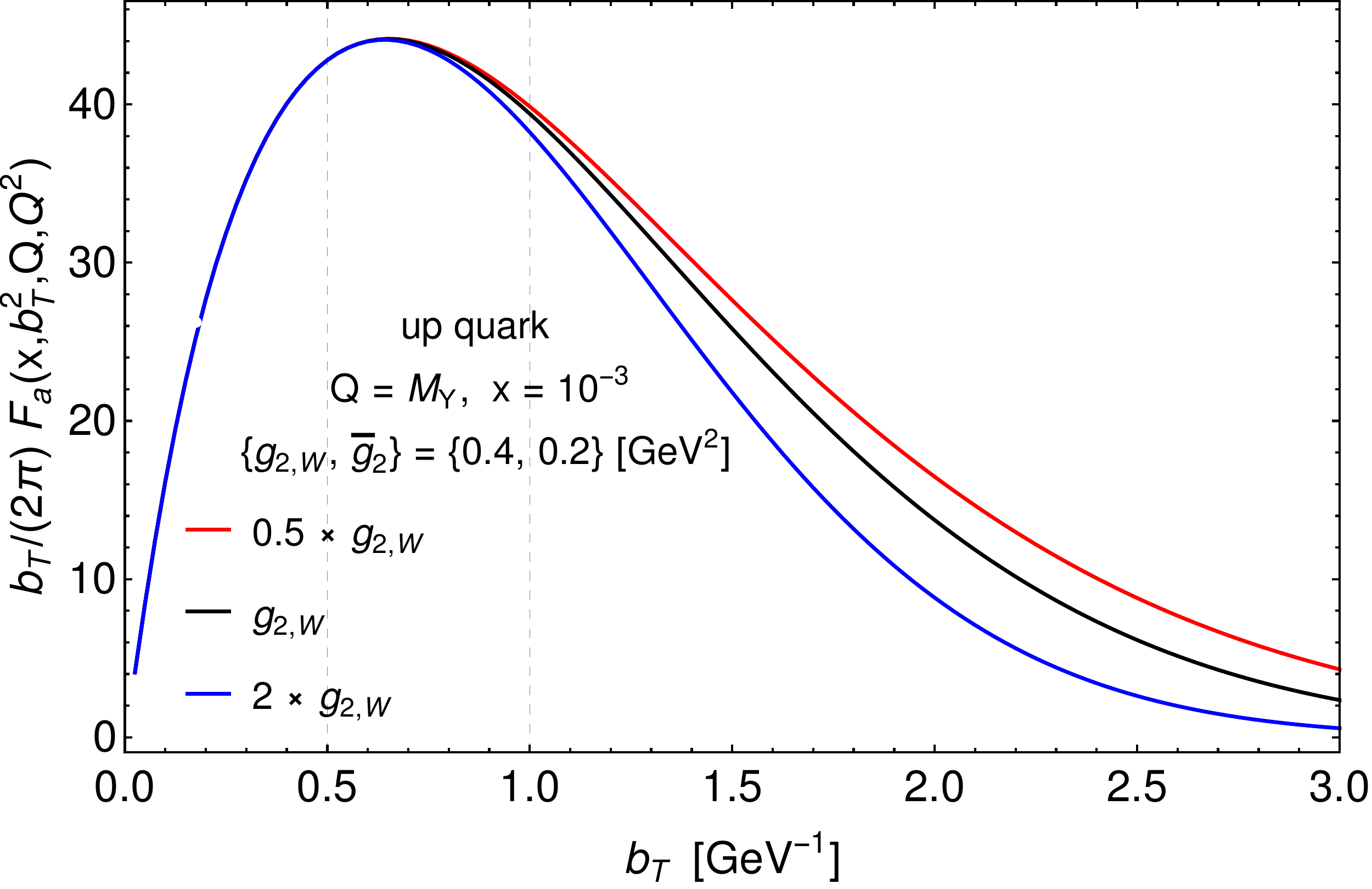}
&\hspace{0.001cm}
&
\includegraphics[width=0.475\textwidth]{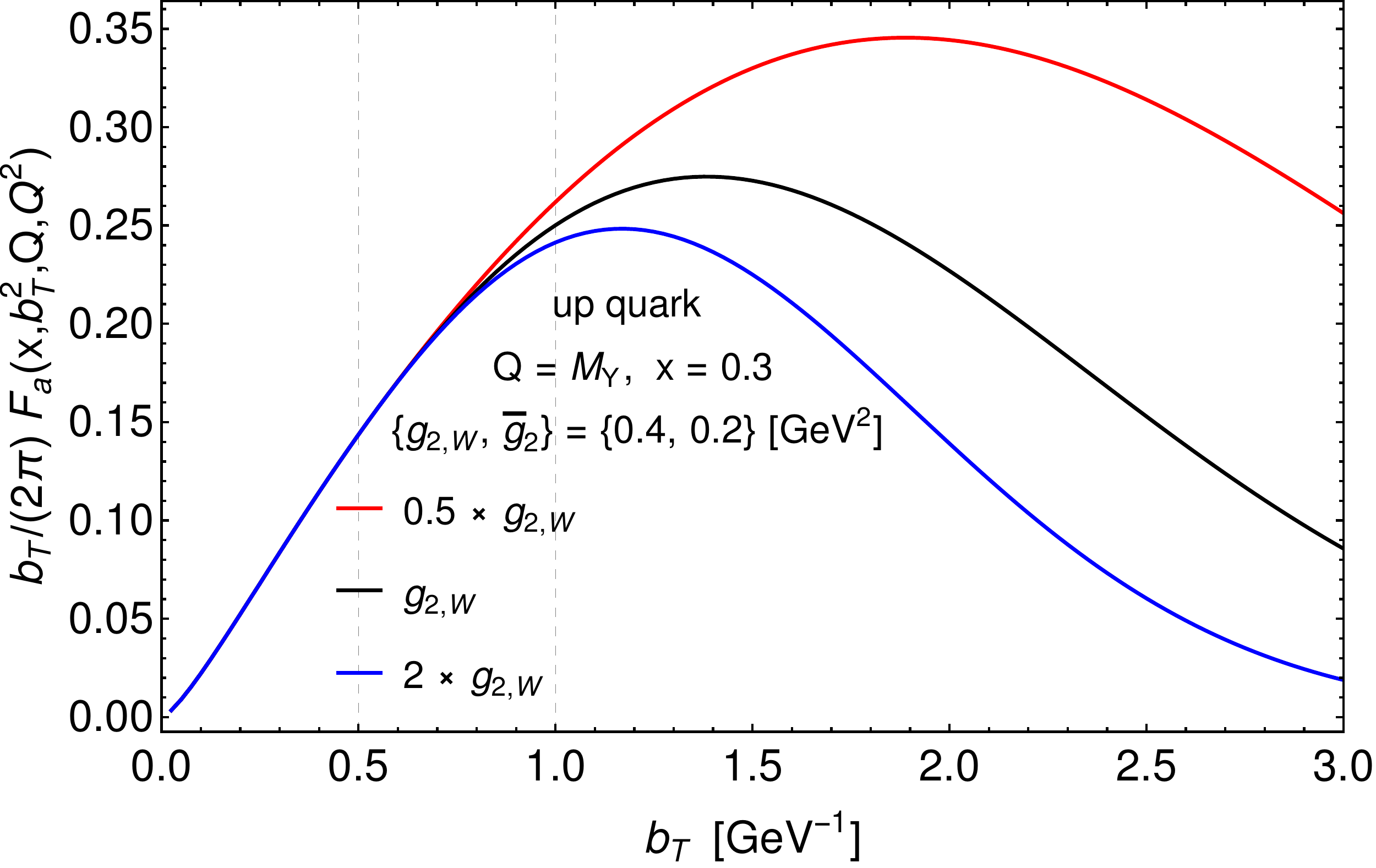}
\\
(c) && (d)
\end{tabular}
\caption{The integrand 
$b_T F_a(x,b_T^2; Q, Q^2)/2\pi$ at NNLL/NNLO for an up quark at (a) $Q = M_Z$ and $x=10^{-3}$, (b) $Q = M_Z$ and $x=0.3$, (c) $Q = M_{\Upsilon}$ and $x=10^{-3}$, and (d) $Q = M_{\Upsilon}$ and $x=0.3$. The non-perturbative corrections (specified by the values of the parameters $g_{2,W}$ and $\overline{g}_2$) have a larger impact on the normalization of the TMD PDF at lower $Q$ and larger $x$.}
\label{f:TMD_up}
\end{figure}

\begin{figure}[h!]
\centering
\begin{tabular}{ccc}
\includegraphics[width=0.475\textwidth]{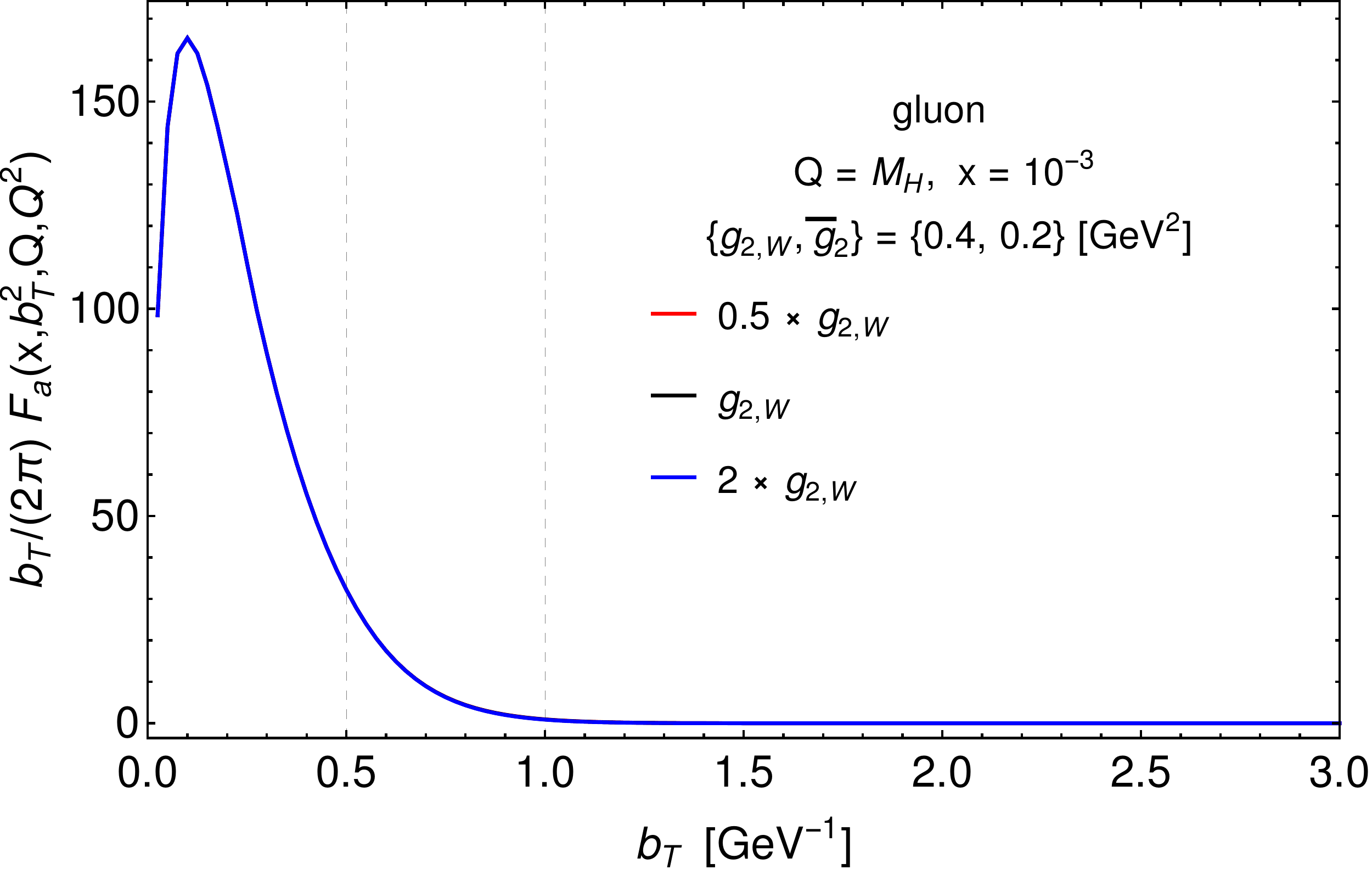}
&\hspace{0.001cm}
&
\includegraphics[width=0.475\textwidth]{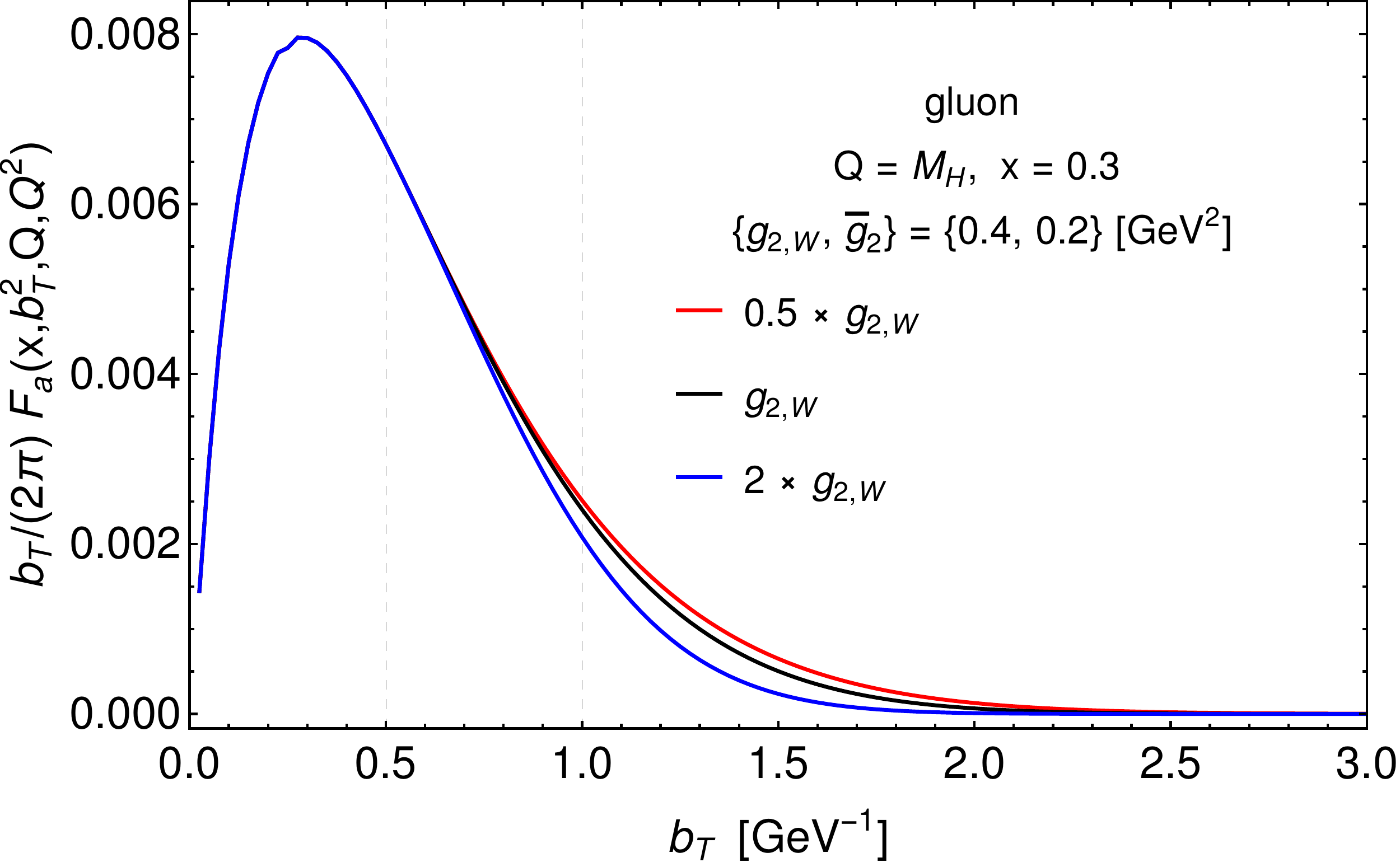}
\\
(a) && (b)
\\ \\ 
\includegraphics[width=0.475\textwidth]{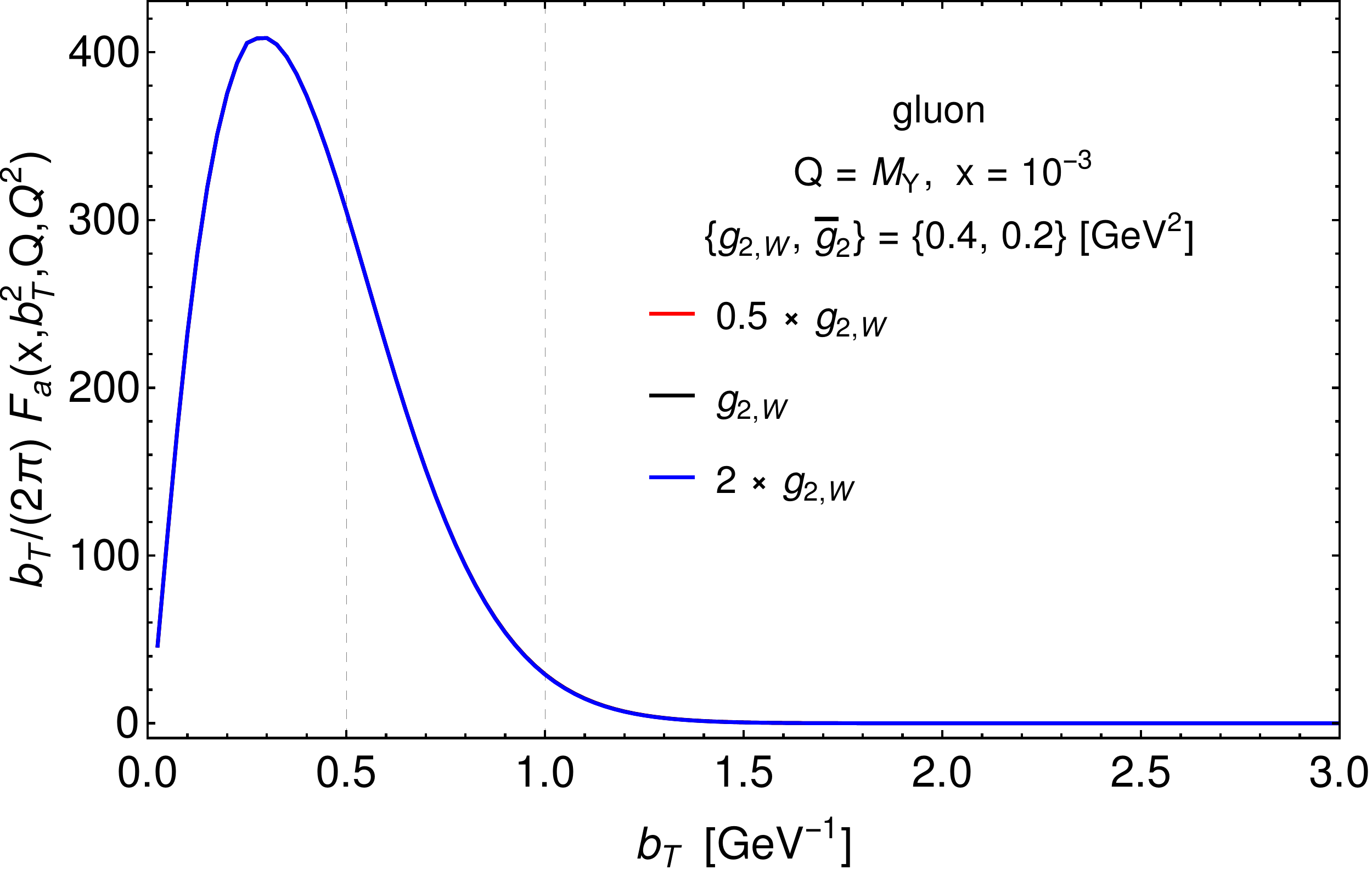}
&\hspace{0.001cm}
&
\includegraphics[width=0.475\textwidth]{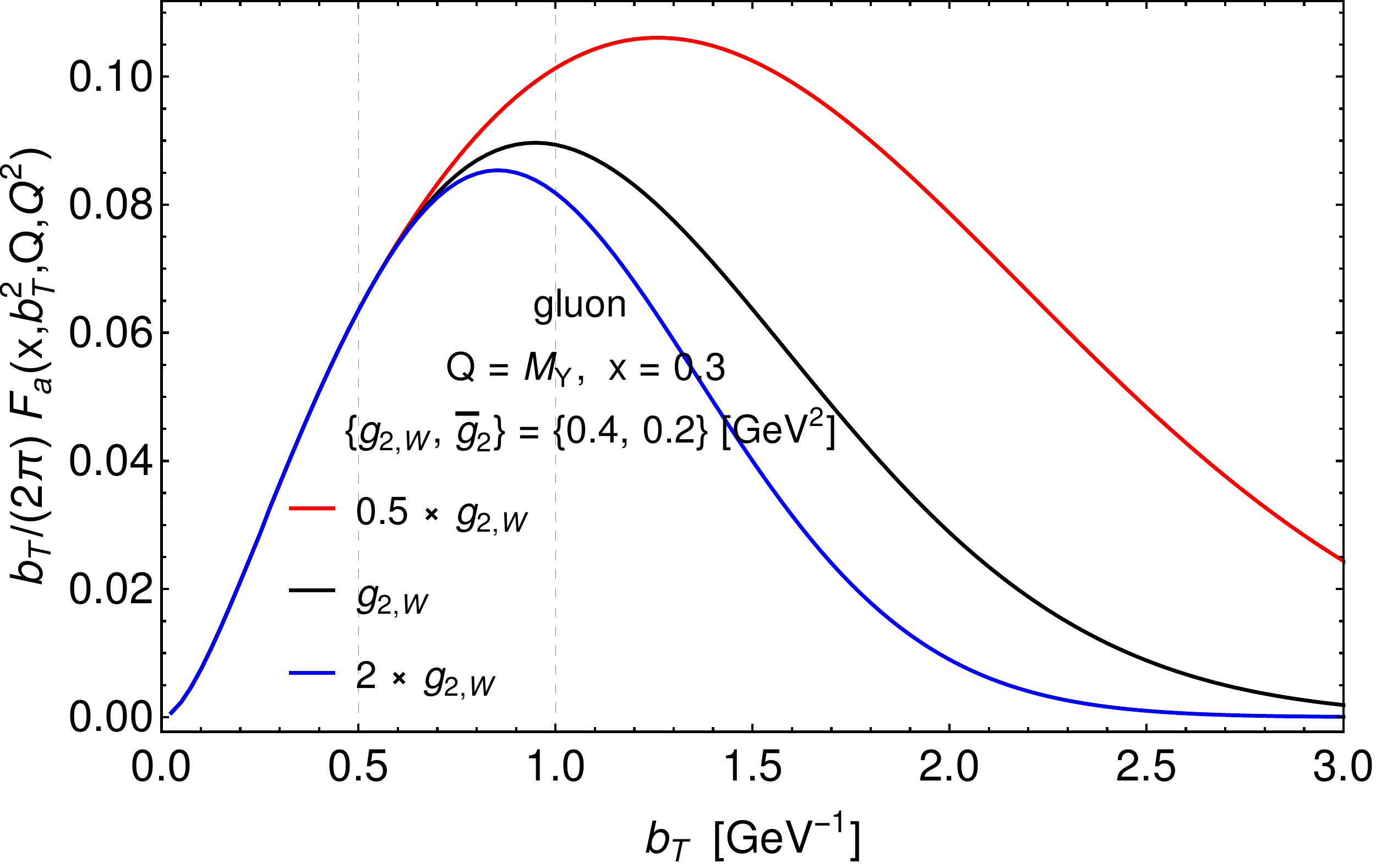}
\\
(c) && (d)
\end{tabular}
\caption{The integrand 
$b_T F_a(x,b_T^2;Q,Q^2)/2\pi$ at NNLL/NNLO for a gluon at (a) $Q = M_{H^0}$ and $x=10^{-3}$, (b) $Q = M_{H^0}$ and $x=0.3$, (c) $Q = M_{\Upsilon}$ and $x=10^{-3}$, and (d) $Q = M_{\Upsilon}$ and $x=0.3$. As for the quark case, the non-perturbative corrections have a sizable impact on the normalization of the TMD PDF at lower $Q$ and larger $x$, even if the impact is less significant with respect to the quark case presented in Fig.~\ref{f:TMD_up}.}
\label{f:TMD_gluon}
\end{figure}

In Fig.~\ref{f:TMD_up} the behavior of the $b_T$-space integrand is displayed for an up quark at $Q=\{M_Z,~M_\Upsilon\}$ and $x=\{10^{-3},~0.3\}$. On the other hand, in Fig.~\ref{f:TMD_gluon} the same quantity is presented for a gluon at $Q=\{M_{H^0}, M_\Upsilon\}$ and $x=\{10^{-3}, 0.3\}$.  
In these figures, it is possible to identify three distinct regions: 
(I) $b_T \lesssim \bmax = 0.5$ GeV$^{-1}$, 
(II) $\bmax \lesssim b_T \lesssim 1$ GeV$^{-1}$, and 
(III) $b_T \gtrsim 1$ GeV$^{-1}$. 
In region I, the integrands are completely determined by the perturbative calculation, see also Eq.~\eqref{e:piecewise_TMD}. Note that by construction this region is not affected at all by the details of the model at large $b_T$. The value of $\bmax= 0.5$ GeV$^{-1}$ is marked with a vertical dashed line in Figs.~\ref{f:TMD_up} and~\ref{f:TMD_gluon}. 
Region II is a transition region from the perturbative to non-perturbative region. Since we require the TMD PDF to be smooth at $b_T=\bmax$, the parameters $(\alpha, g_1)$ in the extrapolation function $R_a^{\rm NP}$ shape the integrand in this region. 
Finally region III is dominated by the physics beyond the leading power/twist QCD perturbative calculations and non-perturbative, 
and the values of the parameters $(g_{2}, \bar{g}_2)$ which quantify the strength of the power corrections would mainly determine the behavior of the integrand. 
Naturally, if the area under region III is very small, the TMD PDF $F_a(x,k_T^2; Q, Q^2)$ in the momentum space will be dominated by the perturbative contribution,
up to the knowledge of 1D PDFs as indicated in Eq.~\eqref{e:input_TMDPDF}. 
On the contrary, if such an area is very large, the TMD PDF in the momentum space will be very sensitive to the non-perturbative contributions. 

In Fig.~\ref{f:TMD_up} and~\ref{f:TMD_gluon}, we fix $\bar{g}_2 = 0.2$ GeV$^2$, and vary $g_{2,W}$ by a factor of 2 up and down from its best fit value $0.4$ GeV$^2$. 
As one can see clearly from Fig.~\ref{f:TMD_up}, for the small-$x$ and large-$Q$ region ($x=10^{-3}$ and $Q=M_Z$), the non-perturbative contribution from the large $b_T$ region $b_T\gtrsim 1$ GeV$^{-1}$ is moderate. But, at the same time, we find that in this region changing $g_{2,W}$ by a factor of 2 leads to minor changes in the integrand, as can be seen from the difference in red and blue curves. This suggests that our derived extrapolation function $R_a^{\rm NP}$ is mainly determined by $g_1$ and $\alpha$ and thus be very good in characterizing the non-perturbative contributions in the large-$b_T$ region. 

For the case of gluons (Fig.~\ref{f:TMD_gluon}), the regions I and II dominantly determine the large-$b_T$ behavior of the integrand at small $x=10^{-3}$, for both values of $Q=\{M_{H^0}, M_{\Upsilon}\}$, while the non-perturbative contribution from the large $b_T\gtrsim 1$~GeV$^{-1}$ to the integrand becomes very small. At large $x = 0.3$, instead, the power corrections have a mild impact at the $H^0$ mass scale and a large impact at the $\Upsilon$ mass scale. Once again, this shows that the value of both the hard scale $Q$ and of the collinear momentum fraction $x$ play an important role in determining the relevance of the large-$b_T$ input in a TMD PDF.

\subsection{Impact of power corrections}
\label{ss:impact_g2}

Let us now study the impact of the power corrections: the dynamical power correction as controlled by $g_2$ and the intrinsic power correction described by ${\bar g}_2$, combined in the parameter $g_2^\prime(Q)$ (see Eq.~\eqref{e:g2p_def}). To quantify the impact of these power corrections on the normalization of $F_a(x,k_T=0;Q, Q^2)$, we study the following ratio:
\begin{equation}
\label{e:Rpc_ratio}
R_{pc}(x,Q\, ;\,g_2^\prime) = \frac{F_a(x,k_T = 0; Q, Q^2)|_{g_2^\prime(M_W)}}{F_a(x,k_T = 0; Q, Q^2)|_{g_{2,W}}} \ ,
\end{equation}
where the intrinsic power correction is fixed to $\bar{g}_2 = 0.2$ GeV$^2$, and $g_{2,W} = 0.4$ GeV$^2$. 
This ratio $R_{pc}$ allows one to focus on the impact of the power corrections only. In Sec.~\ref{ss:impact_extr_pc}, instead, we will focus on the role of the overall extrapolation term.  

We consider $g_2^\prime(M_W) = 2 g_{2,W}$, $g_2^\prime(M_W) = g_{2,W}$, and $g_2^\prime(M_W) = g_{2,W}/2$, which correspond, respectively, to the blue, the black, and the red curves in Figs.~\ref{f:TMD_up} and~\ref{f:TMD_gluon}. 
In Tab.~\ref{t:Rg2_1dm3} and Tab.~\ref{t:Rg2_3dm1} we present the values of the $R_{pc}$ ratio for $x=10^{-3}$ and $x=0.3$, respectively, choosing three different values of $Q$.
Fixing $x$, the impact of the power corrections is generally larger at lower $Q$, which means that the TMD PDF is increasingly affected by the non-perturbative corrections at low energies. 
Viceversa, at fixed $Q$ the impact of the power corrections is more relevant at larger $x$, which means that in the large-$x$ region TMD distributions are affected by potentially large non-perturbative effects. 
At small $x$ (Tab.~\ref{t:Rg2_1dm3}), by changing $Q$ from $M_Z$ to $M_\Upsilon$, the impact of the power corrections on the quark TMD PDF increases by $4-5\%$, whereas at large $x$ (Tab.~\ref{t:Rg2_3dm1}) the increase in the quark case ranges from $12\%$ to $65\%$ for the same change in $Q$. 
Keeping the value of $x$ and $Q$ fixed, the power corrections are less relevant for the gluon, since its TMD PDF is peaked at a lower value of $b_T$ with respect to the quark case (e.g. compare the $Q=M_\Upsilon$ cases in Fig.~\ref{f:TMD_up} and Fig.~\ref{f:TMD_gluon}), due to the Casimir rescaling in the evolution kernel. At small $x$ (Tab.~\ref{t:Rg2_1dm3}), the impact of power corrections on the gluon TMD PDF is very low and is not affected at all by changing $Q$ from $M_{H^0}$ to $M_\Upsilon$, whereas at large $x$ (Tab.~\ref{t:Rg2_3dm1}) the impact is comparable to the quark case. 
Overall, this is a complementary way to prove that TMDs at large $Q$ and small $x$ regions are perturbatively dominated. 
\begin{table}[htp]
\small
 \centering
\begin{tabular}{|c|c|c|c|}
  \hline
  \multicolumn{1}{|c|}{}&\multicolumn{3}{|c|}{$x = 0.001$} \\
  \hline
  \hline
$R_{pc}(x,Q\, ;\,g_2^\prime)$ & $Q = M_{H^0}$ & $Q = M_Z$ & $Q = M_{\Upsilon}$ \\
\hline
up quark & & 
$\{ \textcolor{red}{+4.4 \%}\, , \textcolor{blue}{-6.4 \%} \}$ & 
$\{ \textcolor{red}{+8.4 \%}\, , \textcolor{blue}{-11.8 \%} \}$ \\             
\hline
gluon 	 & 
$\{ \textcolor{red}{+0.02 \%}\, , \textcolor{blue}{-0.01 \%} \}$ & & 
$\{ \textcolor{red}{+0.02 \%}\, , \textcolor{blue}{-0.01 \%} \}$ \\
\hline
\end{tabular}
\caption{Variations of $F_a(x,k_T=0;Q,Q^2)$ as a function of the strength of the power corrections at $x=0.001$ for different $Q$ values. The reference value (the black curve in Figs.~\ref{f:TMD_up} and~\ref{f:TMD_gluon}) corresponds to $g_{2,W} = 0.4$ GeV$^2$ and $\overline{g}_2=0.2$ GeV$^2$ (see Eq.~\eqref{e:g2p_def}). The blue numbers correspond to $g_2^\prime(M_W) = 2 g_{2,W}$ (the blue curves in Figs.~\ref{f:TMD_up} and~\ref{f:TMD_gluon}) and the red numbers correspond to $g_2^\prime(M_W) = g_{2,W}/2$ (the red curves in Figs.~\ref{f:TMD_up} and~\ref{f:TMD_gluon}).}
\label{t:Rg2_1dm3}
\end{table} 
\begin{table}[htp]
\small
 \centering
\begin{tabular}{|c|c|c|c|}
  \hline
  \multicolumn{1}{|c|}{}&\multicolumn{3}{|c|}{$x = 0.3$} \\
  \hline
  \hline
$R_{pc}(x,Q\, ;\,g_2^\prime)$ & $Q = M_{H^0}$ & $Q = M_Z$ & $Q = M_{\Upsilon}$ \\
\hline
up quark & & 
$\{ \textcolor{red}{+13.1 \%}\, , \textcolor{blue}{-18.6 \%} \}$ & 
$\{ \textcolor{red}{+78.0 \%}\, , \textcolor{blue}{-30.1 \%} \}$ \\             
\hline
gluon 	 & 
$\{ \textcolor{red}{+2.69 \%}\, , \textcolor{blue}{-5.13 \%} \}$ & & 
$\{ \textcolor{red}{+68.4 \%}\, , \textcolor{blue}{-23.4 \%} \}$ \\
\hline
\end{tabular}
\caption{Variations of $F_a(x,k_T=0;Q,Q^2)$ as a function of the strength of the power corrections at $x=0.3$ for different $Q$ values. The reference value (the black curve in Figs.~\ref{f:TMD_up} and~\ref{f:TMD_gluon}) corresponds to $g_{2,W} = 0.4$ GeV$^2$ and $\overline{g}_2=0.2$ GeV$^2$ (see Eq.~\eqref{e:g2p_def}). The blue numbers correspond to $g_2^\prime(M_W) = 2 g_{2,W}$ (the blue curves in Figs.~\ref{f:TMD_up} and~\ref{f:TMD_gluon}) and the red numbers correspond to $g_2^\prime(M_W) = g_{2,W}/2$ (the red curves in Figs.~\ref{f:TMD_up} and~\ref{f:TMD_gluon}).}
\label{t:Rg2_3dm1}
\end{table} 
The choice $k_T=0$ is the simplest case since it implies $J_0(0)=1$ in Eq.~\eqref{eq:F-b-space}. This eliminates any oscillation from the Bessel function, and allows a better insight into the physics of the small-$b_T$ and large-$b_T$ regions. 
When $k_T>0$, the Bessel function $J_0(k_T b_T)$ further suppresses the large-$b_T$ region of the $b_T$ integration.

\subsection{Impact of the complete extrapolation term}
\label{ss:impact_extr_pc}

\begin{figure}[h!]
\centering
\begin{tabular}{ccc}
\includegraphics[width=0.475\textwidth]{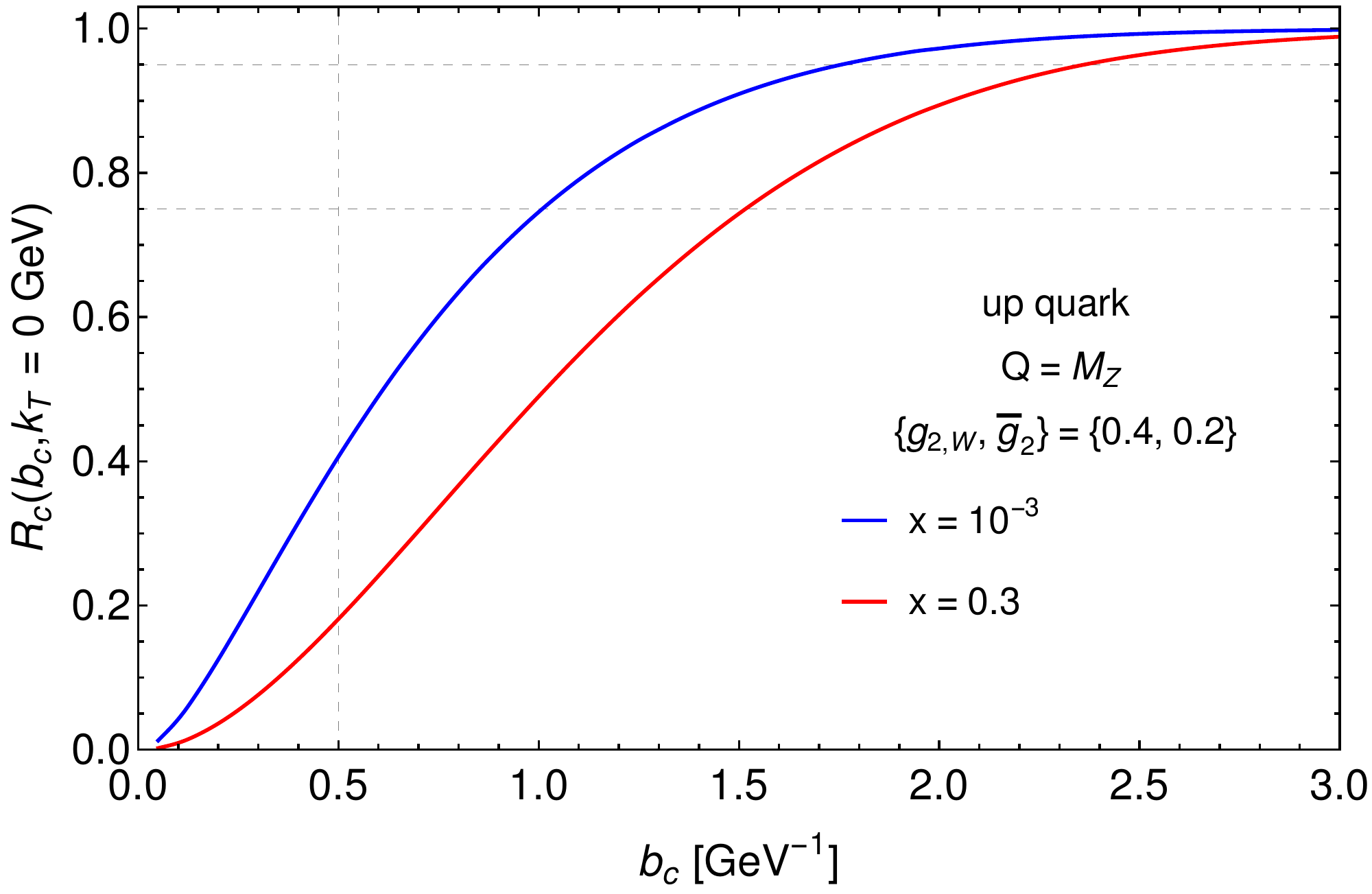}
&\hspace{0.001cm}
&
\includegraphics[width=0.475\textwidth]{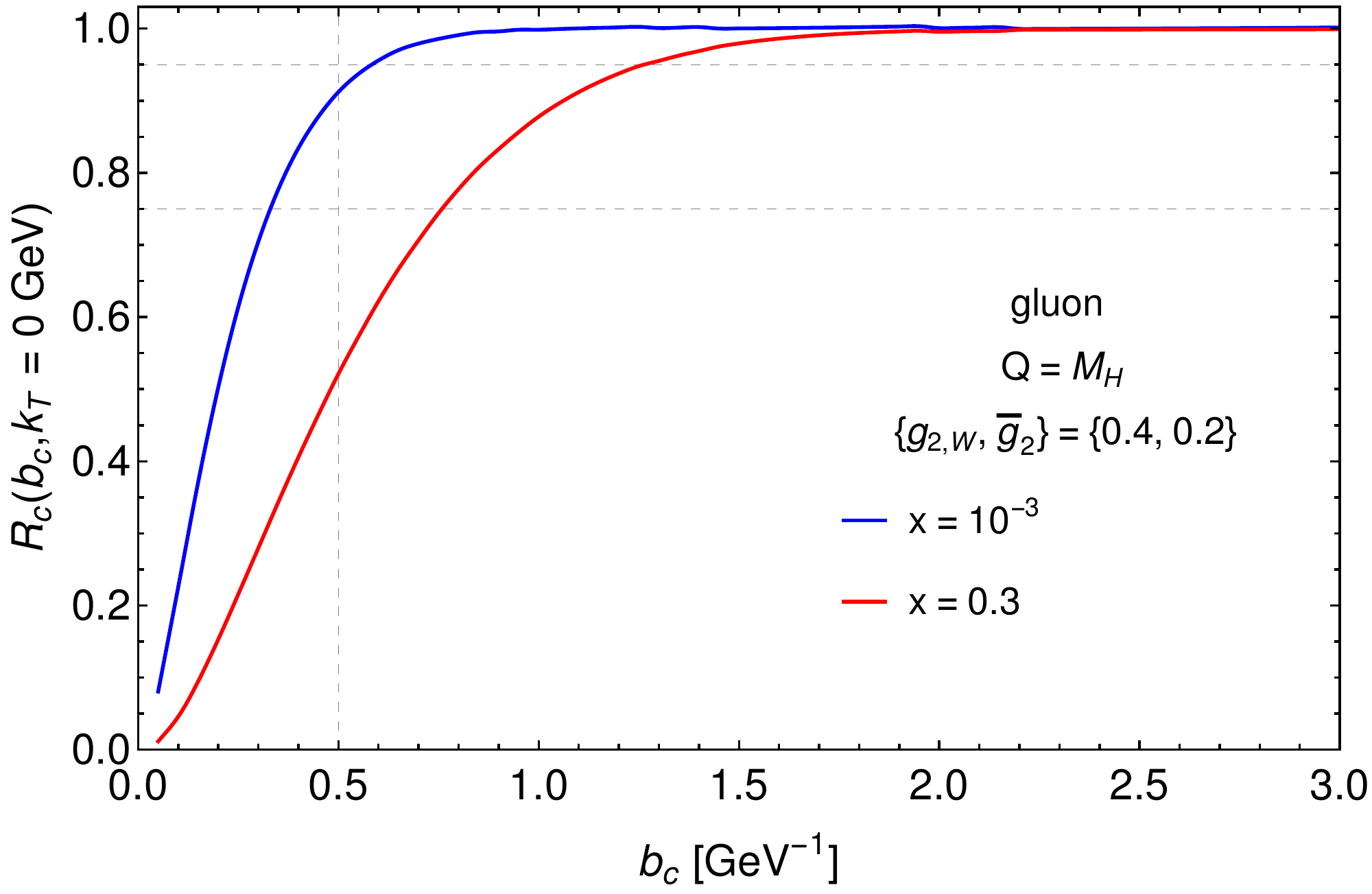}
\\
(a) && (b)
\\ \\ \\ 
\includegraphics[width=0.475\textwidth]{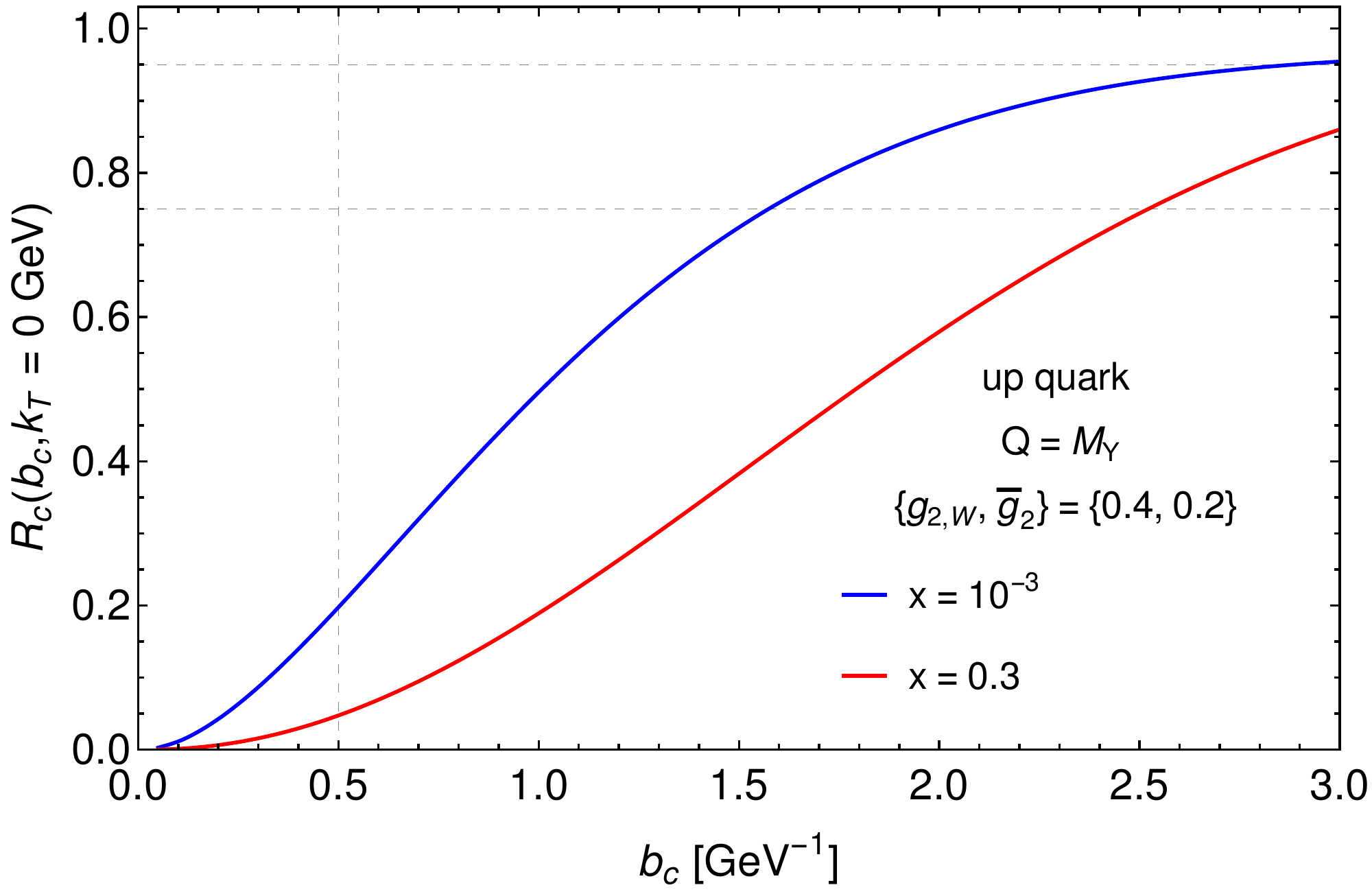}
&\hspace{0.001cm}
&
\includegraphics[width=0.475\textwidth]{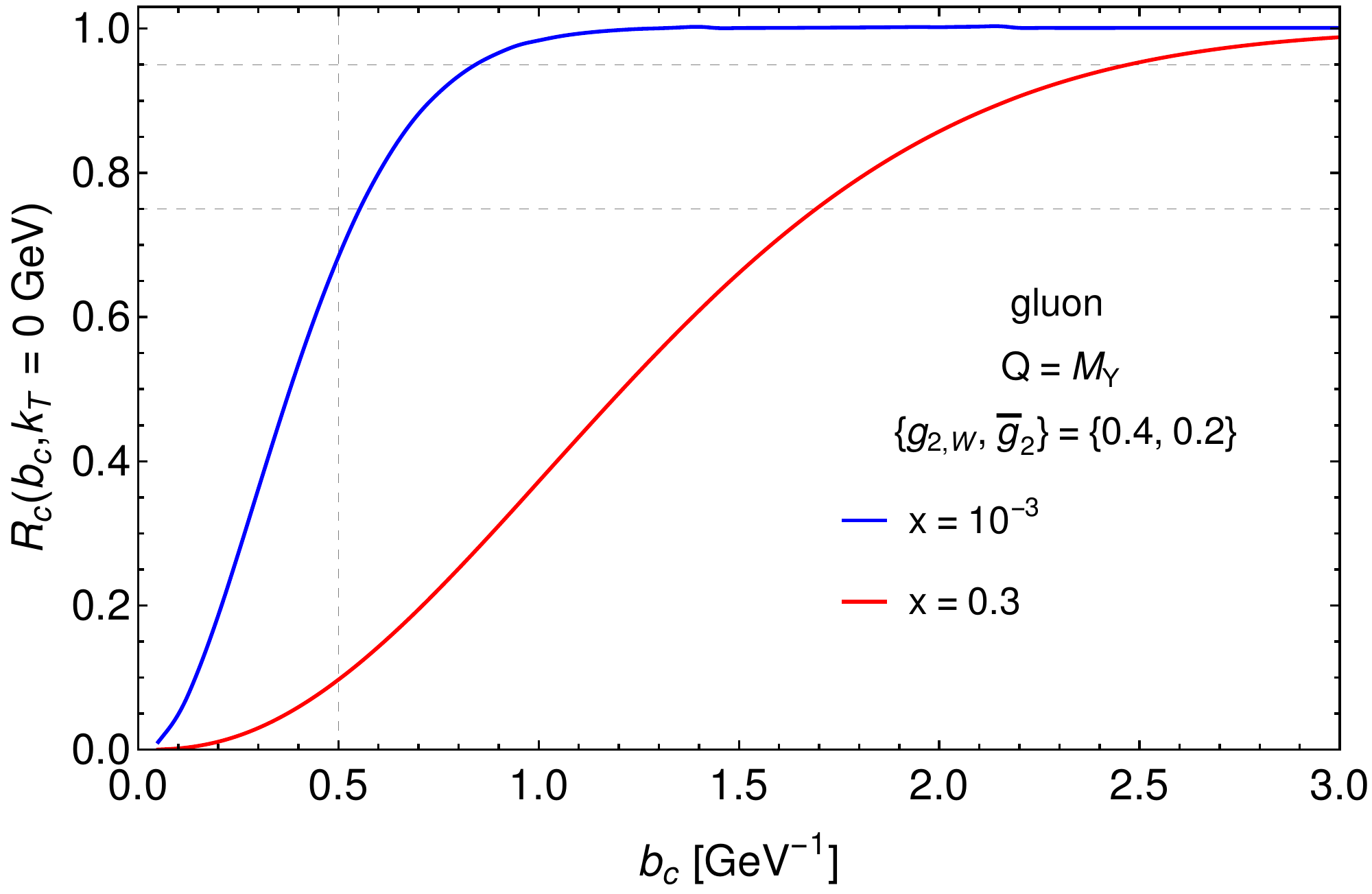}
\\
(c) && (d)
\end{tabular}
\caption{The ratio $R_c(b_c, k_T=0)$ defined in Eq.~\eqref{e:Rc} is plotted as a function of $b_c$ for (a) up quark at $Q=M_Z$, (b) gluon at $Q = M_{H^0}$, (c) up quark at $Q = 9$ GeV, and (d) gluon at $Q = 9$ GeV, respectively. 
The red and blue curves are generated with different choices for the parameters that govern the power corrections (see the legenda).
}
\label{f:Rc}
\end{figure}

Let's introduce a cutoff $b_{c}$ for the upper bound of the $b_T$-space integration in Eq.~\eqref{eq:F-b-space}:
\begin{equation}
\label{e:int_bc}
\omega(b_c, k_T) = \frac{1}{2\pi}\, \int_0^{b_c} db_T\, b_T\, J_0(k_T b_T) F_a(x,b_T^2;Q, Q^2) \ ,
\end{equation}
where $F_a(x,b_T^2; Q, Q^2)$ is defined in Eq.~\eqref{e:piecewise_TMD}. 
To test the influence of the large $b_T$-region on the entire TMD PDF, let's introduce the ratio~\cite{Qiu:2000hf}:
\begin{equation}
\label{e:Rc}
R(b_c, k_T) \equiv \cfrac{\omega(b_c, k_T)}{\omega(b_c\to +\infty, k_T)} \ .
\end{equation}
The ratio $R_c$ represents the fraction of the total integral ($b_c \to +\infty)$ generated by the $0 < b_T < b_c$ region. 
Fig.~\ref{f:Rc} shows the $R_c(b_c, k_T=0)$ ratio for an up quark at $Q = M_Z$ and $Q = M_{\Upsilon}$, and for a gluon at $Q = M_{H^0}$ and $Q = M_{\Upsilon}$. In each panel the ratios computed with $x = 10^{-3}$ and $x=0.3$ are compared. 

Let's consider the value $\bar{b}_c$ such that $R_c(\bar{b}_c, k_T=0) = 0.75$. The latter is highlighted by a horizontal dashed gray line in Fig.~\ref{f:Rc}. 
For an up quark at $Q = M_Z$, $\bar{b}_c \sim 1$ GeV$^{-1}$ at low $x$, whereas at high $x$ one has $\bar{b}_c \sim 1.5$ GeV$^{-1}$. Namely, in order to reproduce $75\%$ of the normalization, a wider portion of the large $b_T$ region is needed at large $x$, where it is thus affected by potentially large non-perturbative corrections. The same trend can be observed in the other three cases too.  
Comparing with Fig.~\ref{f:TMD_up} (a), this also confirms that for an up quark at $Q = M_Z$ and $x = 10^{-3}$ the dominant part of the large-$b_T$ correction is the term proportional to $g_1$ (which is completely determined by imposing the continuity of the first and second derivatives at $b_\text{max}$), whereas at $Q = M_Z$ and $x = 0.3$ also the dynamical and the intrinsic power corrections play a significant role. 

Comparing the panels (a) vs (b) and (c) vs (d) in Fig.~\ref{f:Rc} one sees that, in general, the ratio saturates faster for gluons than for quarks. This is  because the gluon TMD PDF is peaked at lower $b_T$ values with respect to the quark distributions (see Figs.~\ref{f:TMD_up} and~\ref{f:TMD_gluon}) due to the stronger suppression in $b_T$ space generated by the Collins-Soper kernel $K$ and by the UV anomalous dimension $\gamma_F$~\cite{Collins:2011zzd,Echevarria:2012pw,Echevarria:2015uaa}, as already discussed. 
Looking at Fig.~\ref{f:Rc} (a) vs (c) and (b) vs (d) one can see that the effect of lowering the value of the hard scale $Q$ is to increase the sensitivity to the power corrections, both for quarks and gluons and both at low $x$ and large $x$. 
Comparing Fig.~\ref{f:Rc} (b) with Fig.~\ref{f:TMD_gluon} (a) and (b) we can see that for a gluon at $Q = M_{H^0}$ only the term proportional to $g_1$ is relevant to build the 75\% of the total integral. 
A similar argument holds for Fig.~\ref{f:Rc} (d), but comparing with Fig.~\ref{f:TMD_gluon} (c) and (d) we can see that lowering $Q$ the distribution becomes increasingly more sensitive also to the power corrections at large $x$, on top of the $g_1$ term.  
In all cases apart from the gluon at $Q = M_{H^0}$ and $x = 10^{-3}$, comparing with Figs.~\ref{f:TMD_up} and~\ref{f:TMD_gluon} one can see that both the $g_1$ term and the power corrections (namely the overall non-perturbative functions that extrapolates the low $b_T$ behavior into the large $b_T$ region) become relevant to determine the $95\%$ of the TMD PDF at $k_T=0$. 
Eventually, from Fig.~\ref{f:Rc} (b) we determine that for a gluon with $Q = M_{H^0}$ and $x = 10^{-3}$ the fully perturbative region determined by $b_T < \bmax = 0.5$ GeV$^{-1}$ generates the $90\%$ of the TMD PDF at $k_T=0$, whereas at $x = 0.3$ it accounts only for 50\% of the distribution. This shows that, in principle, the transverse momentum distribution of a Higgs boson produced in gluon-gluon fusion in hadronic collisions can receive non-negligible non-perturbative corrections when one of the two collinear momentum fractions is very large~\cite{Echevarria:2012pw}, e.g., for the kinematic region far away from the central-rapidity region at the LHC.  

\section{Cross sections}
\label{s:cross_sections}

In order to compute the transverse momentum distribution of a $Z$ boson or a Higgs boson produced in hadronic collisions we need to calculate the convolution of two TMD PDFs in momentum space. This corresponds to multiplying the two TMD distribution in the $b_T$-space.

\subsection{$Z$-boson}
\label{ss:Z_prod}

For $Z$ production in $pp$ collisions the cross section differential in the transverse momentum $q_T$ and in the rapidity of the produced $Z$ in the low $q_T \ll M_Z$ region reads~\cite{Echevarria:2014xaa,Kang:2012am,Bacchetta:2017gcc}:
\begin{equation}
\label{e:TMDsigma_Z}
\frac{d\sigma^{Z (\to \ell+ \ell^-)}}{dy\, d^2 q_T} = \frac{{\cal H}^Z_0}{2\pi	}\, \sum_q \big( V_q^2 + A_q^2 \big)\, 
\int_0^{+\infty} db_T b_T\, J_0(b_T q_T)\, F_{q/A}(x_A, b_T^2; M_Z, M_Z^2)\, F_{\bar{q}/B}(x_B, b_T^2; M_Z, M_Z^2) \, ,
\end{equation}
where we have neglected the large $q_T$ corrections ${\cal O}(q_T/M_Z)$ to TMD factorization and the corrections ${\cal O}(\Lambda_{\rm QCD}/M_Z)$ to collinear factorization. 
The factors $V_q$ and $A_q$ are the vector and axial couplings respectively of the $Z$ boson to the quark. 
The ${\cal H}_0^Z$ function is the hard function for $Z$-production: 
\begin{equation}
\label{e:hardf_Z}
{\cal H}_0^Z = \sigma_0^{Z (\to \ell+ \ell^-)}\, {\cal H} \, , \ \ \ \ \ \ \ \ \ \sigma_0^{Z (\to \ell+ \ell^-)} = \frac{\sqrt{2}\pi G_F M_Z^2}{s N_c}\, B_R(Z \to \ell^+ \ell^-) \, ,
\end{equation}
where $\sigma_0^Z$ is the leading order term~\cite{Echevarria:2014xaa} and ${\cal H}$ is the hard function for Drell-Yan with the lowest order normalization ${\cal H}^{(0)} = 1$, which we consider at NNLO~\cite{Becher:2008cf}.   
We also adopted the narrow-width approximation, i.e., we neglect contributions for $Q \neq M_Z$. The value of the branching ratio into leptons is $B_R(Z \to \ell^+ \ell^-) = 0.033658$~\cite{Tanabashi:2018oca}. 

As already mentioned, the net effect of multiplying two TMD PDFs in $b_T$ space is that the predictive power for the cross section calculation at a specific value of $x$ and $Q$ is increased with respect to the computation of a single TMD distribution, since the product of two TMDs is peaked at a lower $b_T$ with respect to a single TMD PDF.  

\begin{figure}
\centering
\begin{tabular}{ccc}
\includegraphics[width=0.475\textwidth]{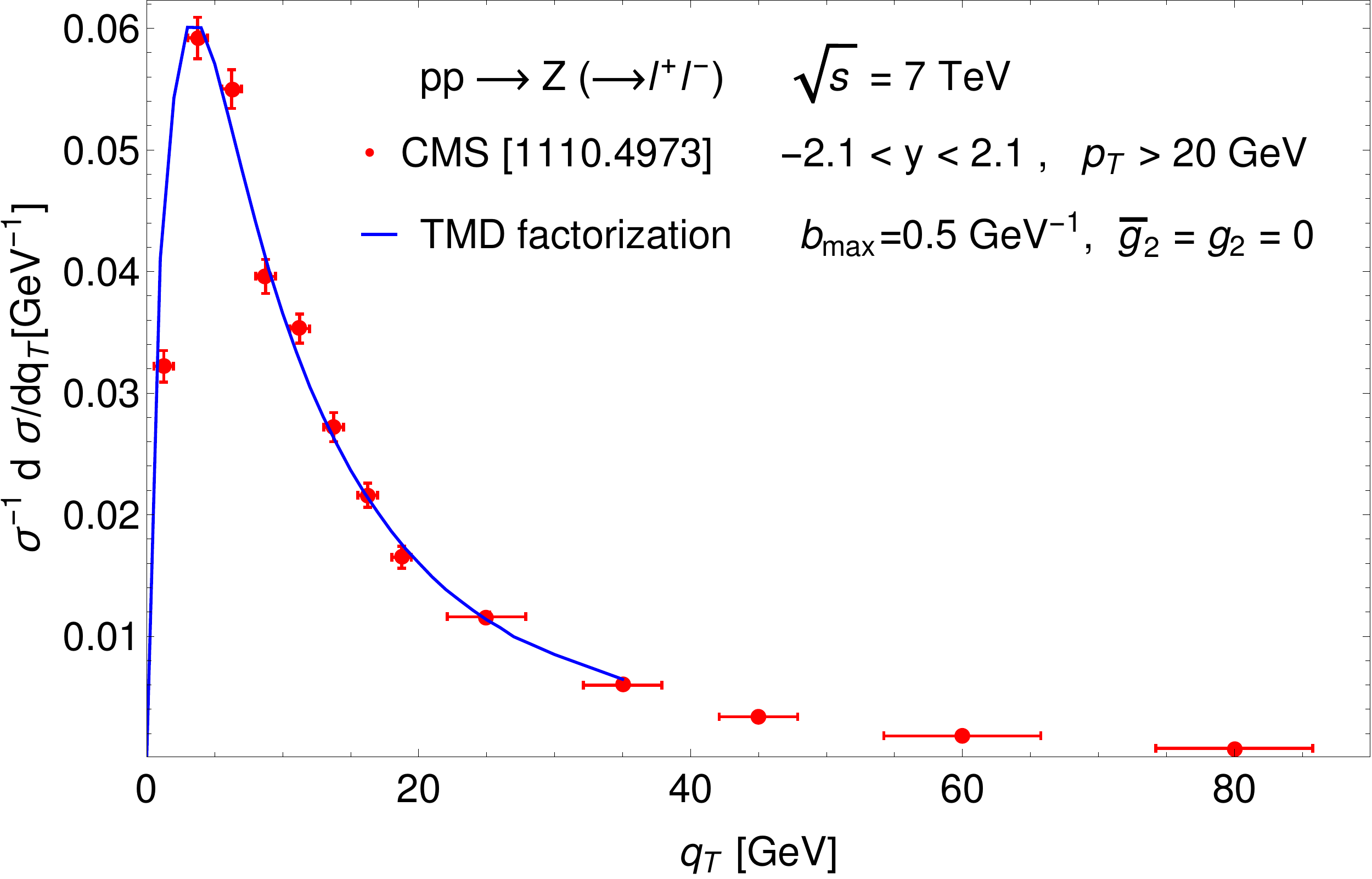}
& \hspace{0.001cm} &
\includegraphics[width=0.475\textwidth]{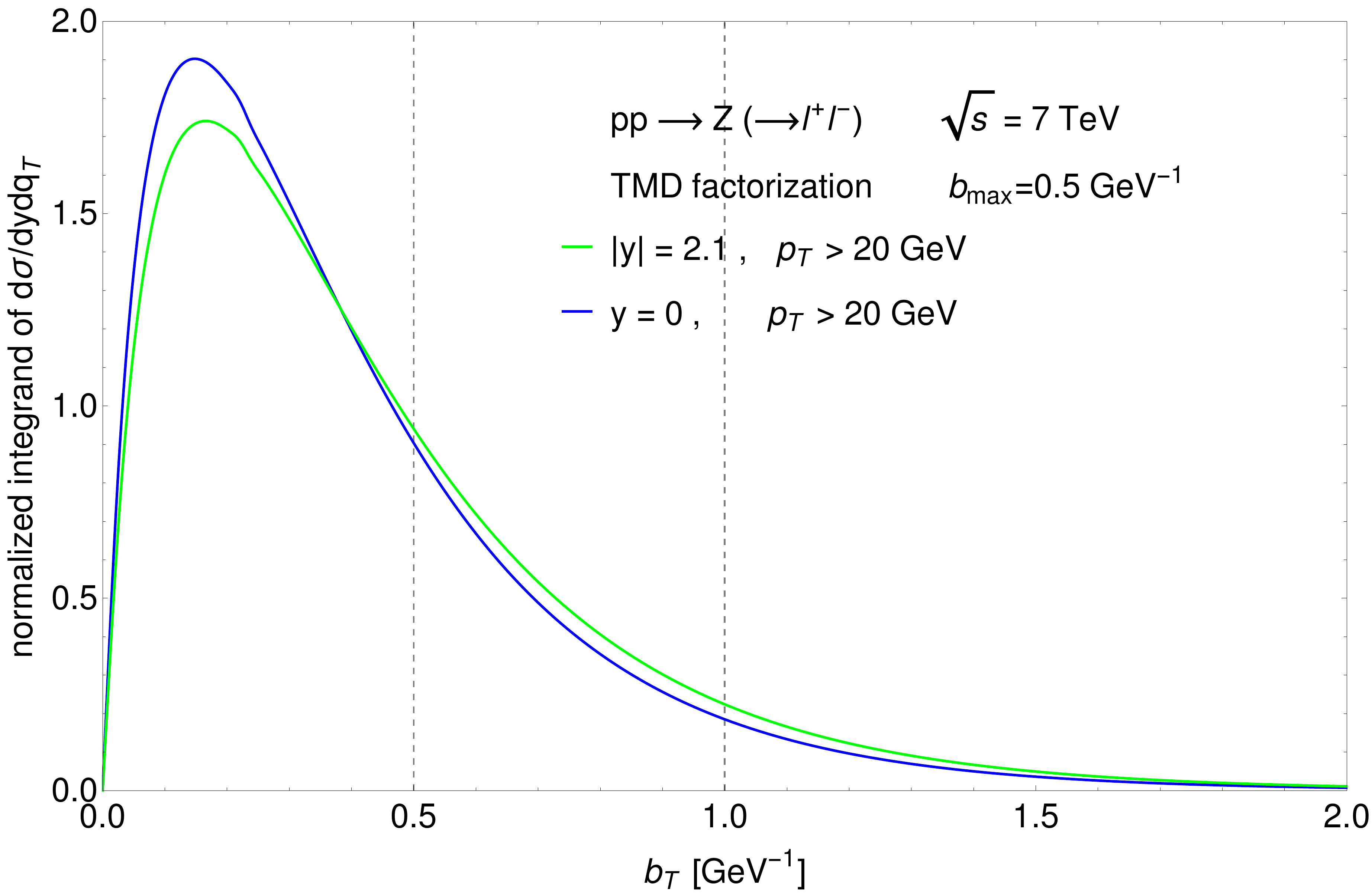}
\\
(a) & & (b)
\\ 
\end{tabular}
\caption{(a) Normalized differential cross section for $Z$ boson production at CMS~\cite{Chatrchyan:2011wt} as a function of transverse momentum $q_T$ and (b) its normalized integrand in $b_T$ space at $q_T=0$. Both for $y=0$ and $|y|= 2.1$ the peak of the integrand and the saddle point of the cross section lie well in the perturbative region ($b_T < 0.5$ GeV$^{-1}$). 
Thus the formalism is predictive and we can describe the data at low $q_T$ just with the perturbative contributions plus the $g_1$-term in the extrapolation function, but without intrinsic or dynamical power corrections. We stress that the theory-data comparison in (a) is not the result of a fit. We evaluated numerically the inclusive cross section $\sigma$ and we find $\sigma = 12.46$ nb.}
\label{f:sigma_Z}
\end{figure}

For example, from Fig.~\ref{f:Rc} (a) we determined that the term $\propto g_1$ in the extrapolation function $R_a^{\rm NP}$, as well as the power corrections play a role in determining the value of the quark TMD PDF at $Q=M_Z$, both at low and high $x$. 
Instead, in Fig.~\ref{f:sigma_Z} (a) we show that we can reproduce the data collected by the CMS experiment at the LHC with $\sqrt{s} = 7$ TeV and central rapidity $-2.1 < y < 2.1$~\cite{Chatrchyan:2011wt} without including any dynamical or intrinsic power correction in the TMD PDF. 
The $g_1$-term in $R_a^{\rm NP}$ is sufficient (and necessary) to capture the behavior of the TMD PDF at large $b_T$ needed to describe the experimental data. No fit to the data has been performed to reproduce the experimental data in Fig.~\ref{f:sigma_Z} (a). 
In Fig.~\ref{f:sigma_Z} (b), instead, the normalized integrand of the differential cross section in $b_T$ space is displayed for $q_T=0$. For the rapidity values $y=\pm 2.1$ and $y=0$ the peak of the integrand lies well in the perturbative region.  
It is also straightforward to check that the peak of $\ln\, ( b_T\, \times $ the cross section integrand$)$, which corresponds to the analogue of the  saddle point for the TMD PDF discussed in Eq.~\eqref{e:sp_equation}, lies at $b_T < 0.5$ GeV$^{-1}$. 
This result is obtained implementing the OPE on the collinear PDFs at small $b_T$ at ${\cal O}(\alpha_s^2)$ and working at NNLL accuracy with the Collins-Soper kernel~$K$ and the UV-anomalous dimension~$\gamma_F$. 

The $x$ range spanned by the data in Fig.~\ref{f:sigma_Z} is $\sim [10^{-3}, 10^{-1}]$ (calculated as $(Q/\sqrt{s})\, e^{\pm y}$). 
We checked that for the data collected at more forward rapidity, where one of the momentum fraction $x$ lies in a large $x$ region (e.g. the one by the LHCb experiment~\cite{Aaij:2015gna,Aaij:2015zlq,Aaij:2016mgv}), the perturbative contribution plus the $g_1$-term alone is not sufficient to correctly describe the data, given also their very high precision. This is consistent with our expectation as the relevance of the non-perturbative contribution increases as $x$ gets larger. Indeed it has been recently shown that the large $b_T$ part of the TMD PDF is relevant if one wants to describe the very precise LHC data at forward rapidity, and it is also important to take into consideration its kinematic dependence~\cite{Scimemi:2017etj,Bertone:2019nxa,Bacchetta:2019sam}. 

Along these lines, we remind that the  predictive power is not an absolute concept, but is always related to the precision of the observable under consideration. For example, there might be extremely precise observables for which the perturbative contributions plus the $g_1$-term alone are not sufficient to capture the correct behavior at relatively large $b_T$ needed to give an accurate description of the quantity considered, even at large $Q$ and small $x$. This is the case, for example, of the $W$ boson mass, whose determination is sensitive also to the intrinsic transverse momentum dependence and its flavor decomposition~\cite{Bacchetta:2018lna,Bozzi:2019vnl}. 

Another interesting information available from Fig.~\ref{f:sigma_Z} is that the TMD cross section given in Eq.~\eqref{e:TMDsigma_Z}, valid in principle at $q_T \ll M_Z$, can accurately describe the data in a range of transverse momenta up to $\sim M_Z/3$, which is comparable to the values quoted in, e.g., Refs.~\cite{Bacchetta:2019sam,Scimemi:2019cmh}. 
The determination of the range of applicability of the TMD formalism depends both on the perturbative accuracy of the calculation and also on the separation of the $b_T$ regions and on the parametrization of the large $b_T$ behavior. 
The determination of the $q_T$ range in which the TMD factorization/approximation describes well the data should be, in principle, combined with the error associated to the TMD factorization~\cite{Echevarria:2018qyi} and can be a useful piece of information in the context of the matching studies~\cite{Collins:2016hqq,Echevarria:2018qyi}.
%

\subsection{Higgs boson}
\label{ss:H0_prod}

In this section we present the calculation for the transverse momentum differential cross section for Higgs boson production from gluon-gluon fusion in $pp$ collisions at $\sqrt{s} = 13$ TeV based on the discussed structure for the TMD PDFs. 
We calculate the cross section in TMD factorization as~\cite{Echevarria:2015uaa}:
\begin{equation}
\label{e:TMDsigma_H0}
\frac{d\sigma^{H^0}}{dy\, d^2 q_T} = \frac{\sigma_0}{2\pi}\, C_t^2 \, \overline{{\cal H}}\,  
\int_0^{+\infty} db_T b_T\, J_0(b_T q_T)\, F_{g/A}(x_A, b_T^2; M_{H^0}, M_{H^0}^2)\, F_{g/B}(x_B, b_T^2; M_{H^0}, M_{H^0}^2) \, ,
\end{equation}
where we have convoluted two gluon TMD PDFs in momentum space. The coefficient $\sigma_0$ is the Born-level cross section, $C_t$ is the coefficient that integrates out the top quark~\cite{Echevarria:2015uaa}, and $\overline{{\cal H}}$ is the hard function for Higgs boson production, with the normalization $\overline{{\cal H}}^{(0)}$=1 in the lowest order. For the analytic expression of these coefficients we refer to Ref.~\cite{Echevarria:2015uaa}. 
The resummation of large logarithms in the cross section is done by evaluating each perturbative coefficient at its natural scale, and evolving them up  to a common scale by using the respective anomalous dimensions~\cite{Echevarria:2015uaa}. 

\begin{figure}
\centering
\begin{tabular}{ccc}
\includegraphics[width=0.475\textwidth]{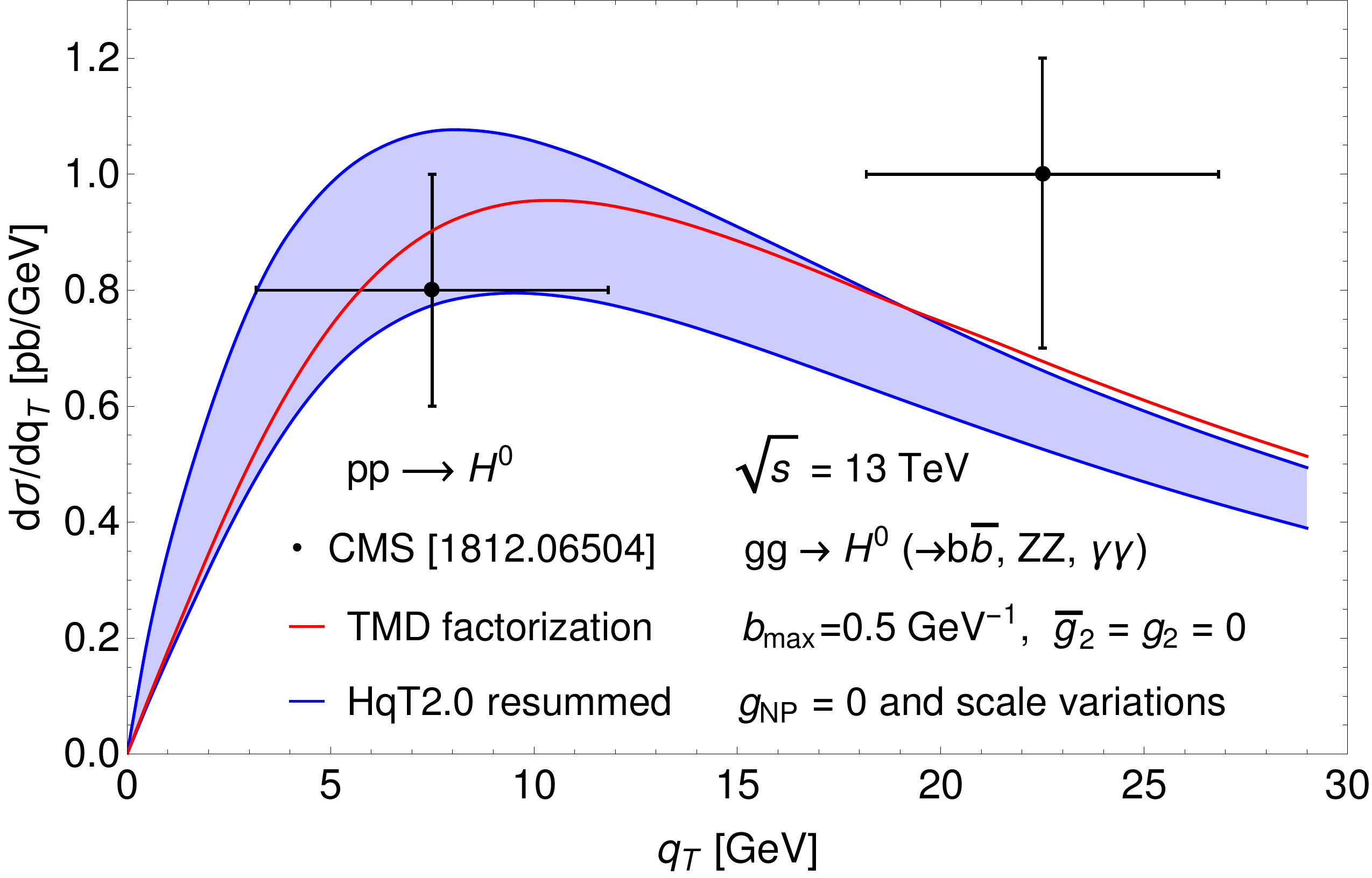}
& \hspace{0.001cm} &
\includegraphics[width=0.475\textwidth]{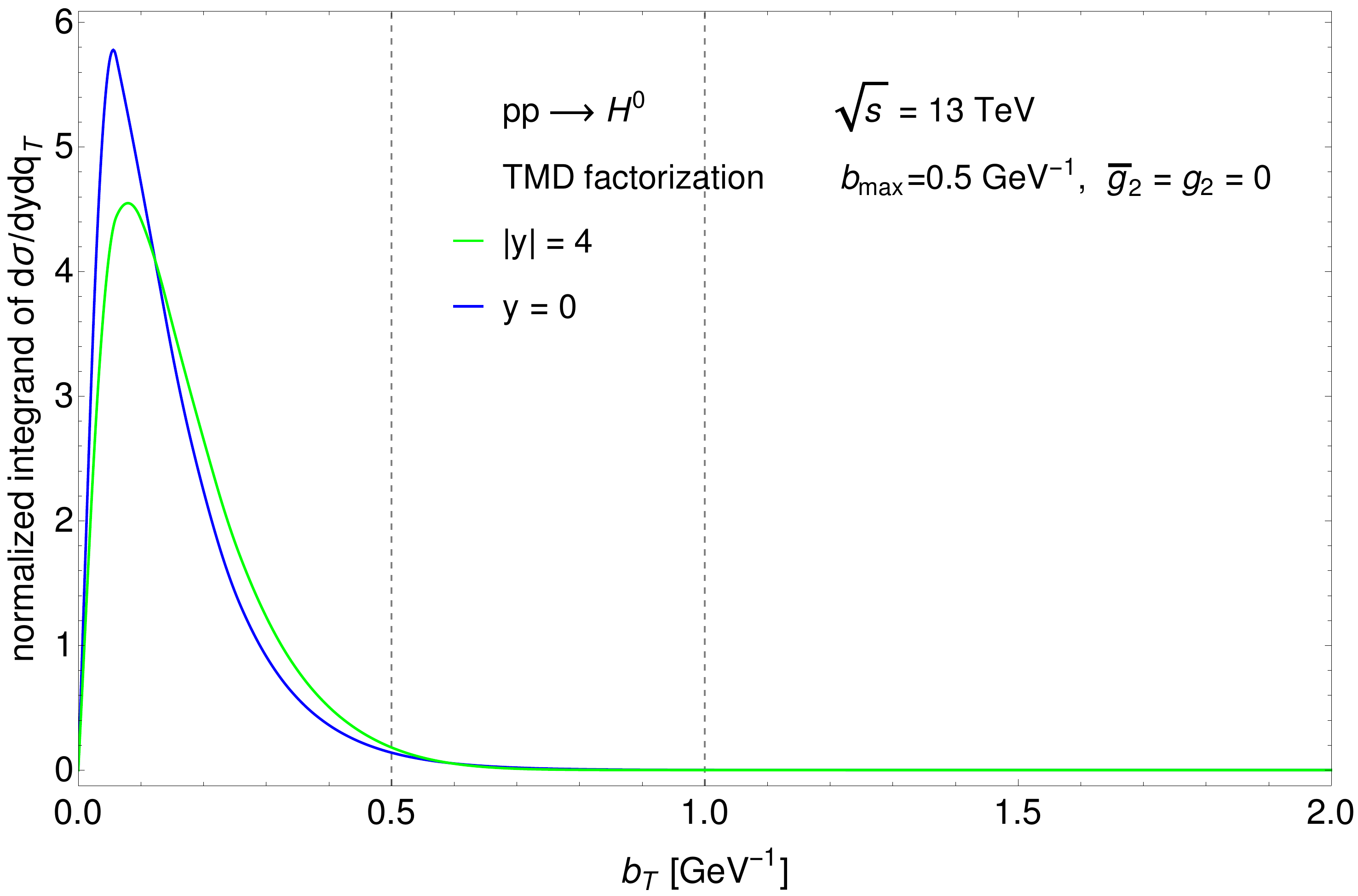}
\\
(a) & & (b)
\\ 
\end{tabular}

\caption{(a) Cross section for Higgs boson production differential with respect to the transverse momentum $q_T$ of the Higgs boson and (b) normalized integrand in $b_T$ space of Eq.~\eqref{e:TMDsigma_H0} at $q_T=0$. 
The data in (a) are from the CMS collaboration~\cite{Sirunyan:2018sgc}. 
The blue band in (a) is built using {\tt HqT2.0} and by varying the perturbative scales around the central value $M_{H^0}$ by a factor of 2. The red curve represents the calculation in TMD factorization with Eq.~\eqref{eq:R-np} assuming an extrapolation to the large $b_T$ region without power corrections ($g_2=\bar{g}_2=0$). The predictions with $g_{2,W} = 0.4, 0.6$ and $\bar{g}_2=0.2$ are identical to the red curve. 
This is because the support of the integrand in (b) is almost entirely in the perturbative region ($b_T < 0.5$ GeV$^{-1}$).}
\label{f:sigma_H0}
\end{figure}

The experimental data available so far to study the Higgs $q_T$ spectrum in the TMD region are affected by very large uncertainties and bin size (see Fig.~\ref{f:sigma_H0} (a)). For this reason, in addition to comparing with the CMS data, we also compare our formalism to another evaluation of the same observable performed in the framework of collinear factorization with transverse momentum resummation. Specifically, we compare to the resummed result available from the public code\footnote{The code is available at \href{http://theory.fi.infn.it/grazzini/codes.html}{http://theory.fi.infn.it/grazzini/codes.html}.} {\tt HqT}~\cite{Bozzi:2005wk,deFlorian:2011xf}. 

Since in this paper we focus only on the unpolarized TMD PDF, we have decided to omit the contribution of the linearly polarized gluons~\cite{Boer:2016xqr} from Eq.~\eqref{e:TMDsigma_H0}. 
Their role in Higgs boson production has been addressed in Ref.~\cite{Boer:2011kf,Boer:2013fca,Boer:2014tka,Echevarria:2015uaa,Chen:2018pzu} and, more recently, in Ref.~\cite{Gutierrez-Reyes:2019rug}. 
Their contribution to the Higgs transverse momentum distribution is known to be of the order of a few percent, depending on the perturbative order and on the implementation of the non-perturbative corrections~\cite{Boer:2011kf,Boer:2014tka,Echevarria:2015uaa}. 
At the phenomenological level the role of the linearly polarized gluons in hadronic collisions is more relevant in the semi-inclusive production of lighter states, such as (pseudo)-scalar quarkonium production at low transverse momentum~\cite{Boer:2012bt,Signori:2016jwo,Lansberg:2017dzg,Echevarria:2019ynx,Scarpa:2019fol}.
%
%
In Fig.~\ref{f:sigma_H0} (a) we compare Eq.~\eqref{e:TMDsigma_H0} at NNLL and NNLO accuracy in TMD factorization with the calculation from {\tt HqT} at the same perturbative accuracy. 
The red curve is the calculation based on the formalism presented in this paper assuming an extrapolation to the large $b_T$ region without power corrections ($g_2=\bar{g}_2=0$). 
{\tt HqT} implements the so-called complex-$b_T$~\cite{Kulesza:2002rh} prescription to separate the small and the large $b_T$ regions. A Gaussian smearing factor in $b_T$ space governed by a single parameter $g_{NP}$ is included to account for the potential non-perturbative effects at large $b_T$. The blue band in Fig.~\ref{f:sigma_H0} (a) has been obtained by setting $g_{NP} = 0$ GeV$^2$ and varying the resummation, renormalization, factorization scales by a factor of 2 around the central value $M_{H^0}$~\cite{Bozzi:2005wk,deFlorian:2011xf}. 
Changing the parameters controlling the non-perturbative corrections in both approaches has a small impact. In particular, the predictions obtained within our formalism using $g_{2,W} = 0.4, 0.6$ and $\bar{g}_2=0.2$ are identical to the red curve in Fig.~\ref{f:sigma_H0} (a). This is because 
the support of the $b_T$-space integrand in Fig.~\ref{f:sigma_H0} (b) is almost entirely in the perturbative region ($b_T < 0.5$ GeV$^{-1}$). 
The two calculations are in good agreement and compatible within the uncertainty band, and the differences (especially for $q_T \gtrsim 20$ GeV) could be due to the different methods employed to separate the small and the large $b_T$ regions.

\section{Summary and outlook}
\label{s:conclusions}

In this paper we have discussed the predictive power of unpolarized transverse momentum dependent parton distribution functions (TMD PDFs, or simply, TMDs), as a function of the light-cone momentum fraction $x$ and of the energy scale $Q$. Such TMD PDFs are essential ingredients in the modern TMD factorization formalism, which generally describes the observables with more than one momentum transfer, such as hadron production in semi-inclusive deep inelastic scattering (SIDIS), and the transverse momentum distribution of vector boson $W/Z$ and $H^0$ production in hadronic collisions. We have determined that the predictive power is maximal in the large $Q$ and small $x$ kinematic region, for example for vector boson production at hadron colliders with $\sqrt{s}$ of the order of the TeV and at central rapidity. In other words, the transverse momentum dependence of the TMDs, probed in this region, is dominated by the leading power and perturbatively calculable contributions from the parton shower in the hard collision, and, therefore, the TMDs in this kinematic region, so as the transverse momentum distributions of the bosons, are 
well predicted 
by the TMD factorization formalism. Outside of this region, the non-perturbative contributions (as represented by the dynamical and intrinsic power corrections in our study) become increasingly relevant, according to the kinematics explored (non-central rapidity, low $Q$, large $x$). Of course this should not be seen as a problem,
rather an advantage for probing the nature of hadron structure. 

We emphasized that the transverse momentum $k_T$-dependence of parton (quark or gluon) TMDs probed with two-scale observables, $Q\gg q_T \gtrsim \Lambda_{\rm QCD}$, in high energy scattering is different from the intrinsic $k_T$-dependence of quarks or gluons inside a bound hadron.  The difference between the measured $k_T$-dependence of an active parton participating in the hard collision and the parton's intrinsic motion is a result of the QCD evolution of the TMDs.  
If the evolution is dominated by the perturbatively calculable kernels at small $b_T$, the observed $k_T$-dependence is effectively generated perturbatively, as pointed out in this paper in the region where $Q$ is large and $x$ is small.  Such measured $k_T$-dependence of the TMDs is not sensitive to the details of non-perturbative hadron structure other than that included in the 1D PDFs, while its predictiveness is critically important for understanding the production of Higgs particles and other relevant observables.  
On the other hand, if the measured $k_T$-dependence and its evolution is dominated by the non-perturbative large $b_T$ region, which corresponds to the large $x$ and/or not too large $Q$ regime as pointed out in this paper, experimental data of such observables could provide the much needed information for extracting the non-perturbative $k_T$-dependence of the TMDs so long as the TMD factorization formalism is valid.  In particular, together with the recent development in extracting the non-perturbative evolution kernels at large-$b_T$ from lattice QCD calculations~\cite{Ebert:2018gzl,Ebert:2019tvc}, we could perform QCD global analysis of such experimental data to extract the intrinsic parton transverse momentum distributions inside a fast moving hadron to shed some lights on the confined motion of quarks and gluons, the fundamental property of hadron structure.

Hadron production at low transverse momentum from SIDIS in the fixed-target mode is probably the configuration where the predictive power from the perturbative contribution alone is the least, and the most sensitive one to the non-perturbative effects~\cite{Signori:2013mda,Anselmino:2013lza,Aidala:2014hva,Bacchetta:2017gcc,Dudek:2012vr}. It is also the most challenging one from the point of view of factorization theorems~\cite{Boglione:2016bph,Moffat:2017sha,Boglione:2019nwk,Liu:2019srj}, given the fairly low value of $Q$ being a couple of GeVs. Vector boson production at RHIC probes a very interesting kinematic region, namely large $Q$, which guarantees that the factorization approximations are well under control, and relatively large $x\sim 0.1$, where the sensitivity to the non-perturbative effects is larger (see Figs.~\ref{f:TMD_up} and~\ref{f:Rc}, where we can see a moderate sensitivity to non-perturbative physics for a quark at $Q = M_{W/Z}$ and large $x$). This could be an optimal kinematic window to study TMD effects, such as the sign change of the Sivers function~\cite{Aschenauer:2015ndk,Kang:2009bp,Boglione:2015zyc}. This also naturally applies to the Drell-Yan measurements at COMPASS~\cite{Aghasyan:2017jop}. 
Another potentially interesting experimental configuration in the same large-$Q$/large-$x$ kinematic region is a fixed-target configuration at the LHC~\cite{Hadjidakis:2018ifr,Kikola:2017hnp}, where several (un)polarized hadron structure measurements could be performed with very high experimental precision, theoretical control on the factorization approximations, and sizable sensitivity to hadron structure effects. 
Last but not least, also the future US-based Electron-Ion Collider~\cite{Accardi:2012qut} will provide new insights in the quest for hadron structure and hadronization and, in particular, on the TMD PDFs and fragmentation functions~\cite{Bacchetta:2015ora,Metz:2016swz,Liu:2018trl,Accardi:2020iqn,Accardi:2019luo,Moffat:2019pci}.  
According to this analysis, a good configuration to probe the quark structure of hadrons in SIDIS at the future EIC could be $\sqrt{s} \sim 100$ GeV for $Q \sim 10$ GeV at central rapidity. At the same $Q$ and rapidity and at higher energies, instead, we would be increasingly sensitive to the perturbative structure of the transverse momentum distributions. 

This investigation can be expanded in several different directions, for example including small-$x$ resummation effects, polarization effects, studying fragmentation functions, and confronting the given parametrization of $R_a^{NP}$ with experimental data from low to high energies. We leave these for future studies.

\begin{acknowledgments}
We thank C. Aidala, L. Gamberg, and E. R. Nocera for stimulating discussions. Z.K. is supported by the National Science Foundation under Grant No.~PHY-1720486.
J.W.Q. and A.S. acknowledge support from U.S. Department of Energy contract DE-AC05-06OR23177, under which Jefferson Science Associates, LLC, manages and operates Jefferson Lab. 
A.S. also acknowledges support from the U.S. Department of Energy, Office of Science, Office of Nuclear Physics, contract no. DE-AC02-06CH11357, and by the European Commission through the Marie Sk\l{}odowska-Curie Action SQuHadron (grant agreement ID: 795475). 
This work is also supported within the framework of the TMD Topical Collaboration. 
\end{acknowledgments}

\bibliographystyle{JHEP}  
\bibliography{TMD_power_v2}


\end{document}

%% file: settings.tex
\pdfoutput=1 

\usepackage[paperwidth=215.9mm,paperheight=279.4mm,centering,hmargin=2cm,vmargin=2.5cm]{geometry}
\usepackage[tbtags]{amsmath}  
\usepackage{tabularx}
\usepackage{amssymb}          
\usepackage{bm}               
\usepackage{graphicx}         
\usepackage[export]{adjustbox}
\usepackage{hhline,multirow}  
\usepackage{dcolumn}          
\usepackage{slashed}
\usepackage{datetime}
\usepackage{color}
\usepackage[dvipsnames]{xcolor}
\usepackage{slashed}
\usepackage{subfloat}

\usepackage{amscd}            
\usepackage{epsfig}
\usepackage{dcolumn}          
\usepackage[dvipsnames]{xcolor}
\usepackage[normalem]{ulem}
\usepackage{mathtools}
\usepackage{comment}
\usepackage[utf8]{inputenc}

\RequirePackage{color}
\RequirePackage[colorlinks=true
,urlcolor=black
,anchorcolor=black
,citecolor=black
,filecolor=black
,linkcolor=black
,menucolor=black
,linktocpage=black
,pdfproducer=medialab
,pdfa=true
]{hyperref}

\def\bea#1\eea{\begin{align}#1\end{align}}
\newcommand{\bmax}{b_{\text{max}}}

\newcommand{\nocontentsline}[3]{}
\newcommand{\tocless}[2]{\bgroup\let\addcontentsline=\nocontentsline#1{#2}\egroup}